\documentclass[a4paper]{article}

\usepackage[margin=1in]{geometry} % full-width

\usepackage{amsmath}
\usepackage{amsthm}
\usepackage{amssymb}

\usepackage[utf8]{inputenc}
\usepackage{hyperref}
\hypersetup{
    unicode,
    pdfauthor={Anass B. El-Yaagoubi, Hernando Ombao, Jean-Marc Freyermuth},
    pdftitle={A Robust Topological Framework for Detecting Regime Changes in Multi-Trial Experiments with Application to Predictive Maintenance},
    pdfsubject={Real time detection of regime change time series data},
    pdfkeywords={Structural Breaks, Locally Stationary Time Series, Online Detection, Time Series Analysis, Change Point Detection, Non-Stationary Data},
    pdfproducer={LaTeX},
    pdfcreator={pdflatex}
}

\usepackage[authoryear, round]{natbib}
\theoremstyle{plain}
\newtheorem{theorem}{Theorem}

\newtheorem{assumption}{Assumption}[section]

\theoremstyle{definition}
\newtheorem{definition}{Definition}[section]

\newtheorem{remark}[theorem]{Remark}

\usepackage{graphicx, color}
\graphicspath{{fig/}}

\newcommand{\bit}{\begin{itemize}}
\newcommand{\eit}{\end{itemize}}
\newcommand{\CAL}[1]{\mathcal{#1}}

\newcommand{\beq}{\begin{equation}}
\newcommand{\eeq}{\end{equation}}

\newcommand{\bea}{\begin{eqnarray}}
\newcommand{\eea}{\end{eqnarray}}
\newcommand{\beas}{\begin{eqnarray*}}
\newcommand{\eeas}{\end{eqnarray*}}

\newcommand{\noi}{\noindent}

\newcommand{\var}{\mathrm{Var}}

\newcommand{\cov}{\mathrm{Cov}}

\usepackage{algorithm, algpseudocode} % use algorithm and algorithmicx for typesetting algorithms
\usepackage{mathrsfs} % for \mathscr command

\title{A Robust Topological Framework for Detecting Regime Changes in Multi-Trial Experiments with Application to Predictive Maintenance}
\author{Anass B. El-Yaagoubi$^1$, Jean-Marc Freyermuth$^2$, Hernando Ombao$^1$}
\date{
    $^1$KAUST, Statistics Program \\ \texttt{\{anass.bourakna, hernando.ombao\}@kaust.edu.sa} \\
    \vspace{3mm}
    $^2$Aix Marseille Univ, CNRS, I2M, Marseille, France. \\ \texttt{jean-marc.freyermuth@univ-amu.fr} \\
    \today
}

\begin{document}
\maketitle
	
\begin{abstract}
    We present a general and flexible framework for detecting regime changes in complex, non-stationary data across multi-trial experiments. Traditional change point detection methods focus on identifying abrupt changes within a single time series (single trial), targeting shifts in statistical properties such as the mean, variance, and spectrum over time within that sole trial. In contrast, our approach considers changes occurring across trials, accommodating changes that may arise within individual trials due to experimental inconsistencies, such as varying delays or event duration. By leveraging diverse metrics to analyze time-frequency characteristics specifically changes in the spectrum and spectrograms and incorporating topological features, our method offers a comprehensive framework for detecting such variations. Our approach can handle different scenarios with various statistical assumptions, including varying levels of stationarity within and across trials, making our framework highly adaptable. We validate our approach through simulations using time-varying autoregressive processes that exhibit different regime changes. Our results demonstrate the effectiveness of detecting changes across trials under diverse conditions. Furthermore, we demonstrate the practical functionality of our method by analyzing the predictive maintenance of the NASA bearing dataset. By examining the time-frequency aspects of vibration signals measured by accelerometers, our approach successfully identifies failures in different bearings, highlighting its potential for early fault detection in mechanical systems.
\end{abstract}


\noindent\textbf{Mathematics Subject Classification (MSC 2020):} 62M10, 62R40, 55N31, 62P30.

\newpage

\section{Introduction}
\label{sec:intro}

Understanding non-stationary phenomena and detecting changes in time series data is crucial across various fields, including economics, climate science, neuroscience and engineering \citep{DYNAMIC_BRAIN_PROCESSES, CP_ENVIRONMENT, CP_CRYPTOS, NON_STATIONARY_BRAIN, CP_LSTS_MONETARY_POLICY}. For example, 
changes in functional connectivity between brain regions can change over time (both in the short-run or long-run), leading to regime changes that may reflect learning progression and shifts in cognitive processes \citep{DYNAMIC_BRAIN_PROCESSES}. Similarly, in climate science, rainfall patterns can vary not just in timing (e.g., occurring earlier vs later in the season) but also in intensity and distribution where such regime changes can affect ecosystems and water resources. 

In this paper, we analyze predictive maintenance where sensors monitoring machinery can help detect regime changes signs of fatigue and early signs of component failure, which is critical for preventing costly breakdowns. Detecting these regime changes has significant practical importance. Early identification can greatly impact outcomes by reducing equipment downtime, preventing costly repairs, and ensuring safety in predictive maintenance. Organizations can implement timely maintenance strategies, leading to substantial cost savings and increased operational efficiency. It is important to note that a regime change is often application-specific. Not every change detected by traditional change point analysis constitutes a regime change. Within an observational window or trial, there might be multiple change points, such as variations in an event's start and end times - but without a change in the event itself. A regime change, however, signifies a fundamental shift in the underlying dynamics or states of the system, indicating a change in the nature of the event itself.

Often, traditional methods for detecting change points focus on analyzing the first two moments, such as the mean and auto-covariance (in the time domain) or the spectrum (in the frequency domain). While effective for certain types of changes, these methods often fail to capture more complex features of the data that involve topological aspects. In this study, we propose a novel approach that focuses on the \underline{topological} features of the time-frequency representation of signals using various metrics. This approach is particularly relevant for multi-trial experiments, where significant changes can occur both within a trial and across trials.

We introduce three methods to detect changes in the spectrum and time-varying spectrum and the topological features derived from these quantities using different distance metrics: L1 norm, L2 norm, and the Wasserstein distance. These methods will be investigated under three scenarios: (1) weakly stationary time series within a trial and across trials; (2) weakly stationary time series within a trial but non-stationary across trials; and (3) locally stationary time series within a trial and non-stationary across trials. By considering these different scenarios and utilizing diverse metrics, we demonstrate that our approach is flexible and capable of capturing subtle changes that traditional methods might miss.

The remainder of this paper is organized as follows. In Section 2, we recall the definitions of the spectrum and time-varying spectrum, and introduce the persistent homology (PH) technique based on sublevel set filtrations of 1D and 2D functions. We present the estimation procedures and outline our regime change detection approach under the three scenarios mentioned earlier. In Section 3, we apply our technique to simulated data and discuss the sensitivity of each metric—L1 norm, L2 norm, and Wasserstein distance under various types of changes in the time-frequency properties of the time series. In Section 4, we demonstrate the functionality of our method by applying it to the NASA bearing dataset, where our approach successfully identifies failures in different bearings based on the vibration signals measured by accelerometers.

\section{Methodology}
\label{sec:methodology}

% In this section, our aim is to introduce the tools that will enable us detects regime changes in univariate time series observed over multiple trials.
Let \( X_r(t) \), where \( t = 1, \ldots, T \) and \( r = 1, \ldots, R \), denote the univariate time series observed over \( R \) trials. Our objective is to detect regime changes from a time-frequency perspective across trials. We begin by recalling the spectral methods for analyzing time series data that utilize the spectrum for weakly stationary time series and the time-varying spectrum for locally stationary time series (or non-stationary time series, in general) \citep{PSD_Priestley_1957}. In addition, we also consider the topological properties of these spectral representations by applying persistent homology of the sublevel set filtrations of these 1D and 2D functions, respectively. This will allow us to capture robust topological features that can indicate regime changes across trials.

Change-point detection is a rapidly evolving field focused on developing, analyzing, and applying methods to segment piecewise stationary or locally stationary time series into quasi-weakly-stationary segments. Applications span scalar, multivariate, and functional time series, with changes potentially involving shifts in the mean, variance, correlations, or spectral properties. Both offline and online change-point detection approaches have been extensively studied in the literature; see, for example, \citet{TRUONG2020107299} and \citet{JIAO2023} for functional data applications and \citet{AUE20092254} and \citet{ASTON2012204} for comprehensive reviews of offline methods.

%ADD MORE REFS.

More recently a great deal of attention has been given to time series of 'objects' such as networks, images \citep{Zhang2024}, high-dimensional covariance structure \citep{Avanesov2018,JIAO2023}. These methods often require a careful proposal for representing these objects in an  efficient manner in order to build changepoint detection methods. Most of the time these representations preserve isometric invariance (metric based dimension reduction methods). For complex objects, it might be preferable to preserve homeomorphic / homotopic invariance. In other words, the goal is to have a more global view of 'shapes'. In this context, Topological Data Analysis (TDA) has emerged as a powerful tool for data scientists, offering innovative approaches to understanding complex data structures \citep{Chazal2021}. An introduction to TDA specifically tailored for multivariate time series applications, particularly in analyzing brain signals, is presented in \citet{TDA_MULTIVARIATE_TS_ANASS}. Among the various TDA methods, persistent homology has garnered significant attention, particularly for its application in change point detection, as demonstrated in studies such as \citet{Mancho2020} and \citet{enikeeva2021changepointdetectiondynamicnetworks}. Additionally, a recent paper by \citet{Guillem2} explores different representations of spectral characteristics in non-stationary audio signals, focusing on the extraction of pertinent topological features for subsequent use in topologically-augmented machine learning algorithms.

TDA offers a novel perspective on time series and signal analysis, introducing innovative methods that extend beyond traditional approaches \citep{severskyTimeSeriesTopologicalData2016,barbarossaTopologicalSignalProcessing2020,barbarossaTopologicalSignalProcessing2020a}. While there has been significant progress in machine learning applications using topologically enhanced features, the complexity of developing rigorous statistical models has left room for further exploration of TDA's potential across various tasks.

Less commonly, some methods aim to segment nonstationary processes into nonstationary—but more homogeneous—segments. For instance, \citet{CASINI2024105811} developed a change point detection method that introduces segmented locally stationary processes. Their approach not only identifies abrupt changes in time at specific frequencies but also detects shifts in the smoothness of the spectrum.

% In this paper, we introduce a novel framework for detecting changes in time series data, particularly suited for analyzing multi-trial nonstationary time series. Our approach is especially effective for applications such as EEG data analysis \cite{OMBAO2018,ASTON2012204}.

In this paper, we focus on detecting regime changes, which involve complex dynamics in the nonstationary properties of processes across multiple observations. Specifically, we analyze locally stationary processes and develop methods to identify changes in regimes. Here, a regime refers to variations in the underlying dynamics observed across trials, with a focus on the topological features of spectral characteristics.

This framework is highly relevant to block or multi-trial experiments, where the \underline{objects} of interest capture the micro-states of a system. For instance, in predictive maintenance, these micro-states reflect the operational characteristics or functioning regimes of machinery. Similarly, in cognitive studies, micro-states (e.g., brain dependence networks) can be characterized using time-evolutionary spectral matrices, coherence matrices, and partial directed coherence matrices.

% To model such objects we will recast our problems into a proper topological framework. In particular, the spectral matrices will be treated as curves or surfaces in the Riemannian manifold of Hermitian positive definite (HPD) matrices.
%\cite{Avanesov2018}. 

\subsection{Spectral Analysis}
\label{subsec:time_frequency_analysis}

Time-series data often display frequency-specific patterns, especially when linked to periodic or repetitive events. Fourier analysis has played a crucial role in identifying these frequency-specific properties, offering insights into the dynamics of such data through their oscillatory behavior. In the following subsections, we will examine both stationary and locally stationary time series, using time-frequency analysis to explore their spectral characteristics and underlying dynamics.

\subsubsection{The spectrum}
\label{subsec:power_spectral_density}

\begin{figure}[H]
    \centering
    \includegraphics[width=\textwidth]{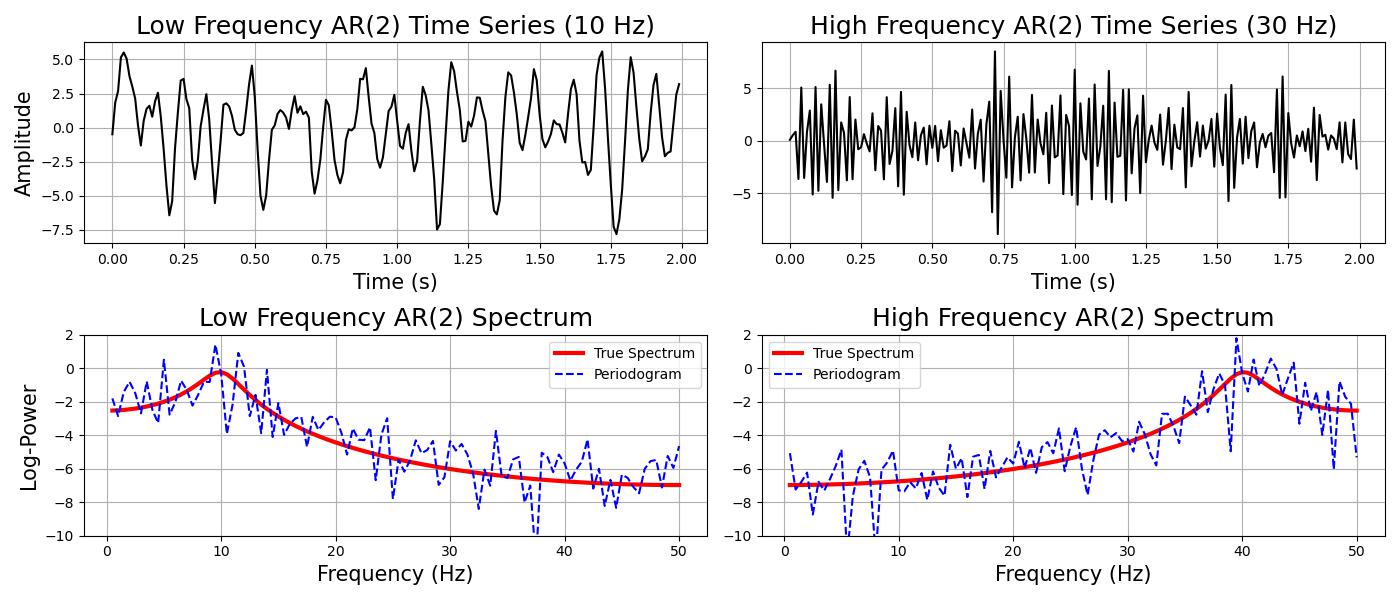}
    \caption{AR(2) process time series and visualization of the spectrum. The top row shows the time series for low-frequency (15 Hz) and high-frequency (35 Hz) AR(2) processes. The bottom row displays the corresponding true spectrum (red) and the periodogram estimator (blue).}
    \label{fig:AR2_PSD}
\end{figure}

Consider a weakly stationary univariate time series \( X(t) \) with \( t = 1, \ldots, T \), characterized by a constant mean \( \mu = \mathbb{E}[X(t)] \) and an autocovariance function \( \gamma(h) = \operatorname{Cov}(X(t), X(t+h)) \) that depends only on the lag \( h \). Moreover, \( X(t) \) admits a Cramér representation, allowing it to be expressed in terms of Fourier waveforms \( e^{i 2 \pi \omega t} \) with random amplitudes \( A(\omega) dZ(\omega) \):
\begin{equation}
    X(t) = \int_{-1/2}^{1/2} A(\omega) e^{i 2 \pi \omega t} \, dZ(\omega).
    \label{eq:cramer}
\end{equation}
here \( A(\omega) \) is the complex-valued transfer function and \( dZ(\omega) \) is complex-valued stochastic process with zero-mean (\( \mathbb{E}[dZ(\omega)] = 0 \)); $\var [dZ(\omega)] = d\omega$; and orthogonal across frequencies so that  $\cov[dZ(\omega), dZ(\omega')] = 0$ for $\omega \neq \omega'$.
Under these assumptions, the spectrum of $X(t)$ is defined as:
\begin{eqnarray}\label{eq:psd_definition}
    S(\omega) = |A(\omega)|^2.
    %\mathrm{Var}\big(A(\omega) dZ(\omega)\big) = |A(\omega)|^2.
\end{eqnarray}
The spectrum describes how the variance of the signal is distributed across the entire frequency range $\omega \in (-1/2, 1/2)$.  
In practice, the spectrum is estimated using the periodogram, as illustrated in Figure~\ref{fig:AR2_PSD}. We define the discrete Fourier transform (DFT) of the time series \( X(\omega) \) and the periodogram \( I(\omega) \) to be, respectively, 
\begin{align}
    X(\omega_k) &= \frac{1}{\sqrt{T}} \sum_{t=1}^{T} X(t) e^{-i 2 \pi \omega_k t},
    \label{eq:dft} \\
    I(\omega_k) &= X(\omega_k) X(\omega_k)^*,
    \label{eq:periodogram}
\end{align}
where \( \omega_k = \frac{k}{T} \) for \( k = 0, 1, \ldots, T-1 \). The Fourier coefficients \( X(\omega_k) \) can be expressed in terms of their real and imaginary parts as: $X(\omega_k) = R_k + i Q_k$, where \( R_k \) and \( Q_k \) are independent and identically distributed (i.i.d.) random variables with zero mean and variance $\frac{S(\omega_k)}{2}$.
%which we will model as normal random variables with mean zero and variance \( \frac{S(\omega_k)}{2} \), i.e., 
%\[
%    R_k, Q_k \sim \mathcal{N}\left(0, %\frac{S(\omega_k)}{2}\right).
%\]
The periodogram \( I(\omega_k) \) is then simply the sum of the squares of the real and imaginary parts: $I(\omega_k) = R_k^2 + Q_k^2$. When the time series length $T$ is sufficiently large, $\frac{2 I(\omega_k)}{S(\omega)} \overset{\cdot}{\sim} \chi_2^2$ (for $0 < \omega_k < \frac{1}{2}$).  
%Since \( A_k \) and \( B_k \) are independent normal random variables, the sum \( A_k^2 + B_k^2 \) converges in distribution to a scaled $\chi_2^2$ distribution with mean equal to the PSD \( S(\omega) \):
%\[
%    \frac{2 I(\omega_k)}{S(\omega)} \overset{d}{\longrightarrow} \chi_2^2.
%\]
The periodogram serves as an estimator of the spectrum. It is asymptotically unbiased, i.e., \( \mathbb{E}[I(\omega_k)] \approx S(\omega) \). However, the periodogram is not mean-squared consistent because its variance does not decrease even as the time series length \( T \) increases. 
%Specifically, the periodogram at each frequency \( \omega_k \) follows an $\chi_2^2$ distribution, resulting in high variance in the estimates, especially at frequencies with high amplitudes.
To obtain a consistent estimator, the periodogram is often smoothed by averaging over neighboring frequency bins. Smoothing reduces the variance of the periodogram, making it consistent as \( T \) increases. Methods for automatic bandwidth selection are given in \cite{PeriodSmootherGCV} and \cite{PeriodSmootherLee}. However, smoothing introduces bias, as the smoothing process can distort the true spectral features around frequencies that display large peaks. Therefore, one must be careful to avoid having a large smoothing window when there is highly localized peak in the periodogram curve. 
%Therefore, there is a trade-off between bias and variance in the estimation of the PSD, and the choice of smoothing technique must balance these factors based on the specific application and data characteristics.

\subsubsection{Spectrogram}
\label{subsec:spectrogram}

\begin{figure}[H]
    \centering
    \includegraphics[width=\textwidth]{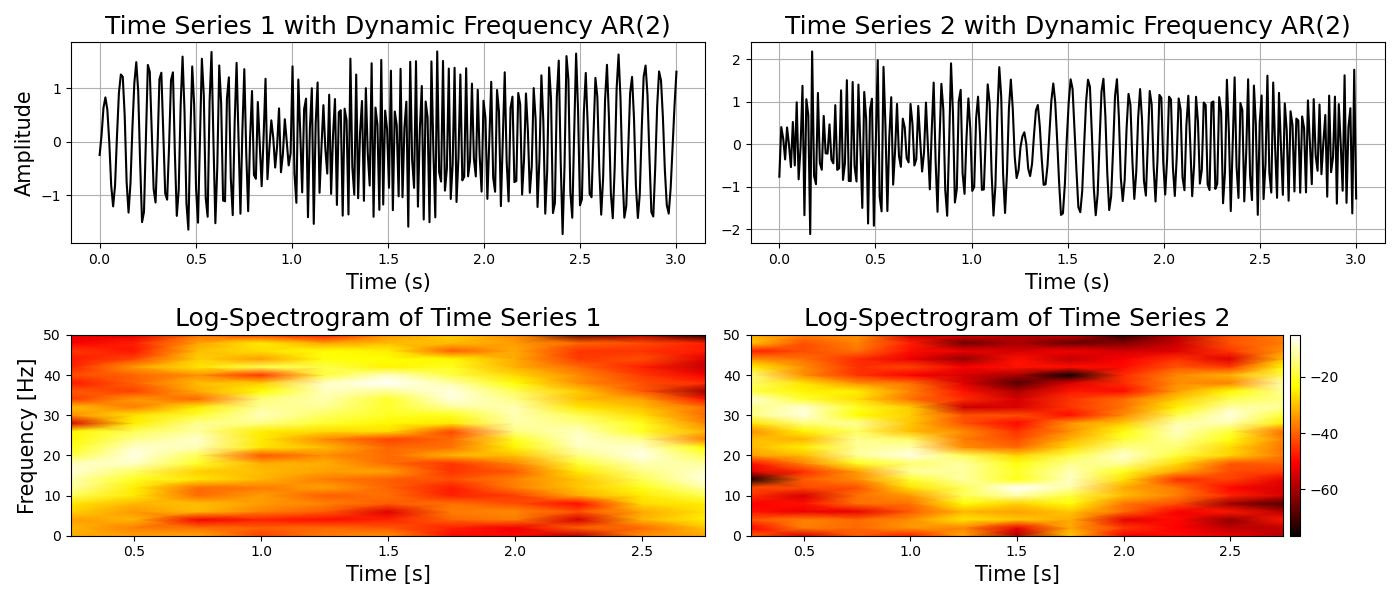}
    \caption{Time-varying AR(2) processes. The left panel shows a frequency pattern that increases from 10 Hz to 40 Hz and then decreases back to 10 Hz, while the right panel shows a pattern that decreases from 40 Hz to 10 Hz and then increases back to 40 Hz. Both patterns evolve over a period of 3 seconds with a sampling rate of 100 Hz.}
    \label{fig:AR2_Spectrogram}
\end{figure}

For non-weakly stationary time series, statistical properties such as the mean, autocovariance (and equivalently, the spectrum) may change over time. In this part, we focus on locally stationary time series, where the spectrum evolves over time in the sense defined in \citet{Dahlhaus}. In this setting, the spectrum alone is insufficient to capture the time-varying spectral characteristics of the signal because it "averages" the distribution of the variance across the entire observation time.  Therefore, it is necessary to estimate the spectrum, locally in time, to account for these temporal variations. The spectrogram, derived from the short-time Fourier transform (STFT), serves this purpose by providing a comprehensive time-frequency representation of the signal. It is computed using the STFT, which analyzes the signal within localized time windows. For a discrete time series \( X(t) \), where \( t = 1, \ldots, T \), the STFT \( \mathcal{S}(\omega_k, t) \) is defined as:
\begin{equation}\label{eq:stft}
\mathcal{S}(\omega_k, t) = \left| \sum_{h = - L}^{L} X(t+h) w(h) e^{-i \omega_k (t+h)} \right|^2,
\end{equation}
where \( w(\cdot) \) is a weight function centered at zero with width \( 2L + 1 \) and normalized such that \( \sum_{h = - L}^{L} w(h)^2 = 1 \). Thus, the spectrogram \( \mathcal{S}(\omega_k, t) \) provides a two-dimensional representation of the signal's power distribution over both time and frequency, as shown in Figure~\ref{fig:AR2_Spectrogram}. The spectrogam captures how the frequency content in a non-stationary signal evolves over time (both in the short-run across time in a trial; and in the long-run over trials in the experiment). 
The estimation of the time-varying spectrum using the spectrogram involves a fundamental trade-off between variance and bias. Utilizing a smaller temporal window size reduces temporal bias (due to local stationarity) by allowing the spectrogram to quickly adapt to temporal changes in the signal, thereby capturing fine-grained temporal variations. However, one must exercise caution with this approach because a small time window has fewer data points which leads to an increase in the variance of the estimator thereby limiting the reliability of the spectral content estimation. Conversely, employing a larger temporal window size decreases variance by averaging over more data points, resulting in a more stable estimate of the spectral properties. But on the flipside, this can introduce temporal bias by potentially smoothing out important transient features and failing to capture rapid changes in the frequency content of the signal.

Similarly, the raw spectrogram \( \mathcal{S}(\omega_k, t) \) can exhibit high variance, particularly in regions where the spectral content changes rapidly. To achieve a more consistent estimate of the time-varying spectrum, the spectrogram can be smoothed across both time and frequency domains. Smoothing effectively reduces the estimator's variance, enhancing its consistency as the sample size \( T \) increases. Nevertheless, it introduces bias by potentially obscuring fine spectral details and transient features. Consequently, the selection of smoothing parameters must carefully balance the need to reduce variance while preserving essential spectral characteristics.

\subsection{Persistent Homology}
\label{subsec:persistent_homology}

In Topological Data Analysis (TDA), we use persistent homology (PH) which is a powerful technique to uncover and analyze the topological features of data across multiple scales \citep{TDA_EDELSBRUNNER,TDA_EDELSBRUNNER_HARER}. This study applies PH to sublevel set filtrations of functions derived from spectral representations of time series data, specifically the smoothed periodograms and the  spectrograms which are, respectively, one-dimensional (1D) and two-dimensional (2D) functional estimator of the spectrum and time-varying spectrum. By focusing on the spatial distribution of local extrema, PH enables us to capture essential topological properties indicative of underlying regime changes in the data.

\begin{figure}[H]
    \centering
    \includegraphics[width=\textwidth]{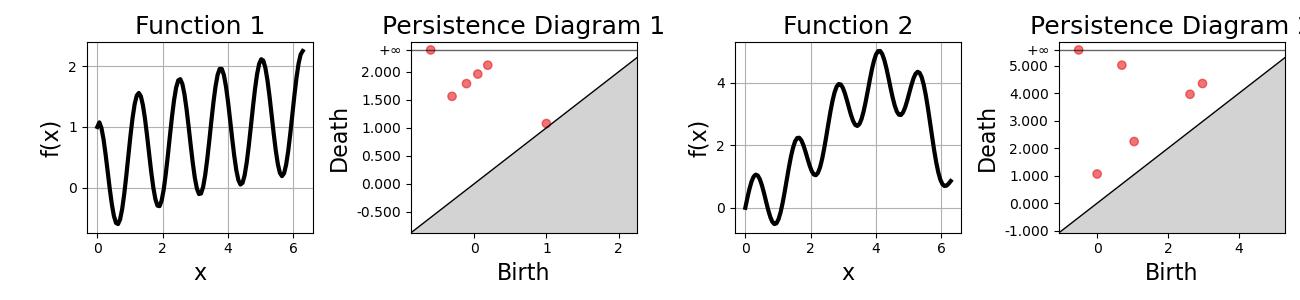}
    \caption{Visualization of two Morse functions and their corresponding persistence diagrams. The first function (left) is a simple Morse function with two periods, while the second function (right) is a more complex Morse function incorporating exponential and cubic components. The persistence diagrams show the birth and death of connected components, reflecting the topological structure of each function.}
    \label{fig:PH_1D_Function_Examples}
\end{figure}

More specifically, PH constructs a filtration (sequence of nested subsets) of the function \( f \) indexed by a scale parameter \( a \). For each value of \( a \), the corresponding sublevel set \( X_a \) is defined as the preimage \( f^{-1}((-\infty, a]) \), which includes all points in the domain of \( f \) where the function value is less than or equal to \( a \). As \( a \) increases, these sublevel sets expand, capturing more of the function’s domain. PH tracks the formation and destruction of topological features such as connected components and one dimensional holes within these sublevel sets, recording the scales at which new components appear (birth times) and disappear (death times). For 1D functions, birth times correspond to local minima, while death times correspond to local maxima. These birth and death pairs provide a comprehensive summary of the function’s topological features across different scales. This information is vital for detecting significant structural changes, such as regime changes, in the underlying data. Figure~\ref{fig:PH_1D_Function_Examples} illustrates this concept using two simple functions. The corresponding persistence diagrams capture the birth and death times of each component within the sublevel set filtration, emphasizing the differences in their structure.

\begin{figure}[H]
    \centering
    \includegraphics[width=\textwidth]{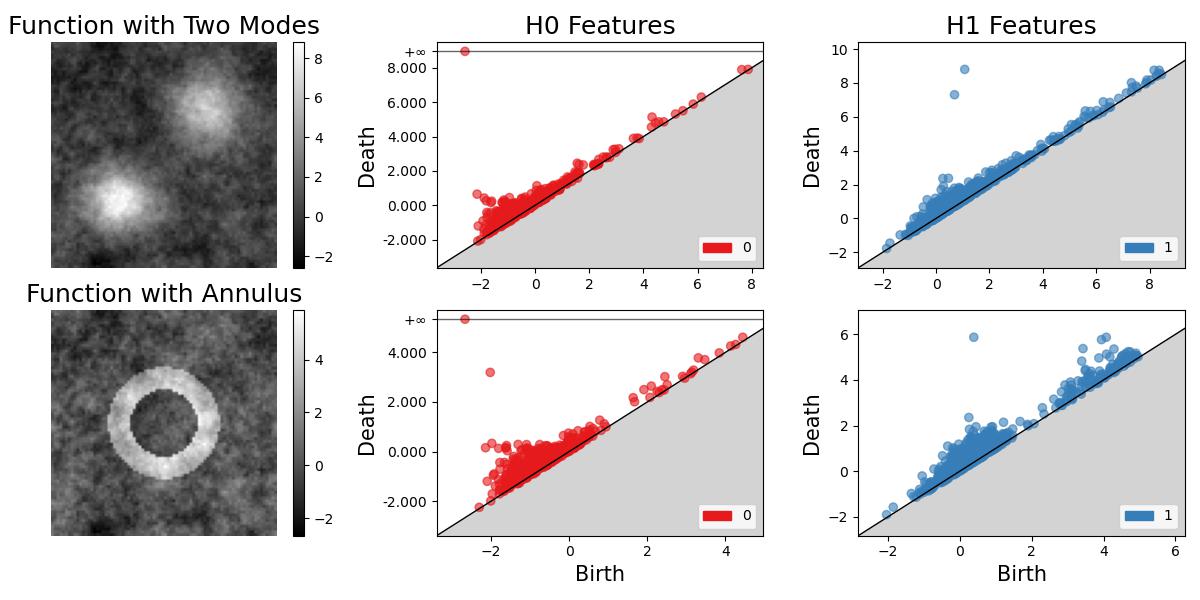}
    \caption{Visualization of two-dimensional functions and their corresponding persistence diagrams. The first row represents a function with two modes, and the second row represents a function with annulus shape. The first column displays the function in gray scale, the second column shows the \( H_0 \) persistence diagram capturing connected components, and the third column shows the \( H_1 \) persistence diagram capturing holes in the function domain.}
    \label{fig:PH_2D_Function_Examples}
\end{figure}

Given a function of dimension 1, i.e., \( f: \mathbb{R} \to \mathbb{R} \), the PH of its sublevel set filtration can only capture features of dimension up to 0. This is reflected in the \( H_0 \) persistence diagram, which displays the connected components of the function's sublevel sets (see the \( 0 \)-dimensional PH diagrams in Figure~\ref{fig:PH_1D_Function_Examples}). However, when the function is of dimension 2, i.e., \( f: \mathbb{R}^2 \to \mathbb{R} \), the PH of its sublevel set filtration can capture features of dimension up to 1. This includes both connected components (\( H_0 \)) and holes (\( H_1 \)), as illustrated in Figure~\ref{fig:PH_2D_Function_Examples}.

In the two-mode example, as the threshold increases, the two peaks are represented as two distinct points in the \( H_1 \) diagram, corresponding to the emergence of two 1D holes. The \( H_0 \) diagram remains empty, as the topology does not include any significant 0D components. However, for the annulus, as the threshold increases, a single prominent point appears in both the \( H_0 \) and \( H_1 \) diagrams. This point in \( H_1 \) represents the 1D hole formed by the annular structure, while the point in \( H_0 \) corresponds to the merging of the components separated by the annulus.

\begin{figure}[H]
    \centering
    \includegraphics[width=\textwidth]{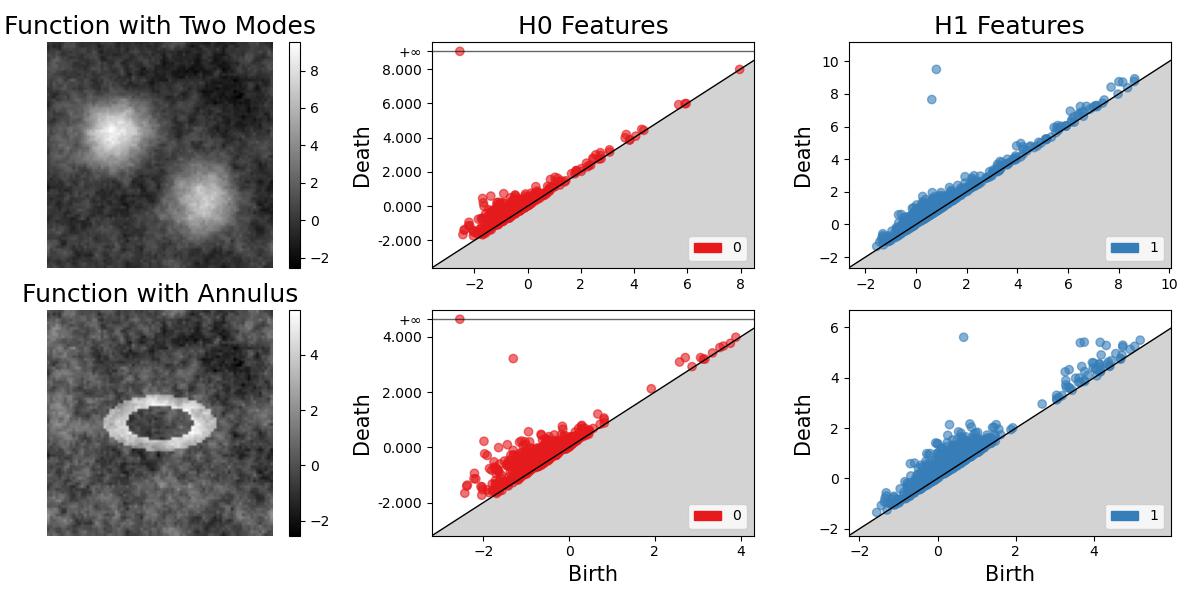}
    \caption{Modified version of Figure \ref{fig:PH_2D_Function_Examples}. The top example demonstrates a function with peaks shifted to different locations, while the bottom example shows horizontal stretching of the original function. Both examples illustrate the invariance of persistence homology under monotonic transformations such translation and stretching.}
    \label{fig:PH_2D_Function_Examples_Bis}
\end{figure}

It is crucial to recognize that PH of sublevel set filtrations remains invariant under monotonic transformations such as shifting or re-scaling. For instance, Figure \ref{fig:PH_2D_Function_Examples_Bis} illustrates this property, where both the translation of peaks and horizontal stretching do not significantly alter the corresponding persistence diagrams. The minor discrepancies observed are due to background noise, not fundamental changes in the topological structure.

This property is particularly important in our context, as it enhances the method’s robustness by avoiding the detection of spurious changes, like shifts or rescaling. Instead, the focus remains on identifying true regime changes rather than simple change points. In multi-trial experiments, it is common for the timing or duration of events to vary across trials, which could lead to high false positive rates in classical methods.

\subsection{Online Change of Regime Detection} \label{subsec:regime_change_detection}

\noi In this subsection, we aim to develop a systematic approach for detecting regime changes in time series data across multiple trials. By leveraging different assumptions regarding the underlying processes, we define specific classes of multi-trial processes that allow us to apply various estimation techniques and statistical tests for regime change detection. 

\begin{definition}[Multi-trial Segmented Locally Stationary Processes]
A centered zero-mean univariate discrete-time multi-trial process $\{ X_t^{(r)},\ t \in \mathbb{N}, r = 1, \ldots, R \}$ is called a multi-trial segmented locally stationary process if it can be represented as:
\begin{equation}
    X_t^{(r)} = \int_{-1/2}^{1/2} A^{(r)}(u, \omega) e^{i2\pi \omega t} dZ^{(r)}(\omega),
    \label{eq:cramer_multitrial_locally_stationary}
\end{equation}
where $A^{(r)}(\omega)$ is a transfer function that depends only on frequency $\omega \in [-1/2,1/2]$ and the trial index $r$. The spectral process $\{Z^{(1)}(\omega), \ldots, Z^{(R)}(\omega), \omega \in [-1/2,1/2]\}$ is a zero-mean stochastic process with orthogonal and cross-orthogonal increments. \label{def:cramer_multitrial_locally_stationary}
\end{definition}

\begin{remark}
Let $\CAL{S}$ denote the function space of time-varying spectral densities. To ensure that the persistent diagrams computed from the sublevel set filtration of these functions are well-defined, several important assumptions must be fulfilled. In particular, we require functions $S \in \CAL{S}$ to be piecewise H\"older continuous. Additionally, certain geometric conditions must be met to prevent the appearance of cusps and corners, and to ensure that no discontinuities occur too close to the boundary of the domain. For details, we refer the reader to \citet{henneuse2024}.   
\end{remark}

For stationary processes, our analysis relies on estimating the spectral density using the periodogram, as detailed in Section \ref{subsec:power_spectral_density}. If the process is stationary across trials, we compare the averaged spectral densities obtained from all previous trials with that of the next trial. However, if the process is not stationary but evolves slowly across trials, the focus shifts to comparing the spectral densities of consecutive trials. For locally stationary processes, we base our analysis on estimating the time-varying spectral density using the spectrogram, as described in Section \ref{subsec:spectrogram}. This approach allows us to account for changes in the spectral characteristics of the process over time.

To develop a robust approach against scaling, shifts, and time warping, we define the equivalence class of spectral characteristics of a process using homological concepts. This method offers a more nuanced comparison than directly using traditional metrics like the $L_1$ or $L_2$ norms between spectral characteristics. The benefits of this approach will be demonstrated in the next section through numerical experiments.

\begin{definition}[Equivalence Class of Spectral Characteristics]
Let $S \in \mathcal{S}$ be the spectral characteristic (either a spectral density or a time-varying spectral density) associated with the observed process $X_t$. The homology equivalence class of $S$, denoted $\left[S\right]$, is defined as:
\[
\left[S\right] = \left\{ \widetilde{S} \in \mathcal{S} \mid \text{Dgm}\left(\text{Filt}(\widetilde{S})\right) \sim \text{Dgm}\left(\text{Filt}(S)\right) \right\},
\]
where $\text{Dgm}\left(\text{Filt}(\cdot)\right)$ denotes the persistence diagram obtained from the sublevel set filtration applied to the smoothed periodogram or spectrogram, and $\sim$ indicates the equivalence relation. Equivalence is determined using the $L_2$-Wasserstein distance, as defined in Definition \ref{def:wasserstein_diagrams}, for a given homology dimension (e.g., \( H_0 \) or \( H_1 \)).
\end{definition}

\begin{definition}[$L_2$-Wasserstein Distance Between Persistence Diagrams]
Let $\text{Dgm}_1$ and $\text{Dgm}_2$ be two persistence diagrams corresponding to the spectral characteristics of two observed processes, for a fixed homology dimension $H_d$ (e.g., \( H_0 \) or \( H_1 \)). The $L_2$-Wasserstein distance between $\text{Dgm}_1$ and $\text{Dgm}_2$ is defined as:
\[
W_2(\text{Dgm}_1, \text{Dgm}_2) = \left( \inf_{\gamma \in \mathcal{B}(\text{Dgm}_1, \text{Dgm}_2)} \sum_{x \in \text{Dgm}_1} \| x - \gamma(x) \|_2^2 \right)^{\frac{1}{2}},
\]
where \(\mathcal{B}(\text{Dgm}_1, \text{Dgm}_2)\) denotes the set of bijections (mappings) between points in \(\text{Dgm}_1\) and \(\text{Dgm}_2\). This distance measures the minimal cost of matching points between the two diagrams, specific to the selected homology dimension, in a way that minimizes the total squared displacement.
\label{def:wasserstein_diagrams}
\end{definition}

\begin{assumption}[Change of Regime Detection]
Assume we observe a sequence of multi-trial processes \(\{X_t^{(1)}, X_t^{(2)}, \ldots\}\) and aim to detect a regime change. There exists a positive integer \(\eta\) such that:
\[
S^{(1)} = \ldots = S^{(\eta)} \neq S^{(\eta+1)} = S^{(\eta+2)} = \ldots
\]
\end{assumption}

\noindent In this context, we define the null and alternative hypotheses at trial \(r\) as follows:

\begin{eqnarray}
&& H_0: S^{(l_0)} = \ldots = S^{(r+1)},\\ %\quad l=l_0,\ldots,r,\\
&& H_1: S^{(l_0)} = \ldots = S^{(r)} \neq S^{(r+1)}.% \quad l=l_0,\ldots,r,
\end{eqnarray}

\begin{assumption}[Homological Change of Regime Detection]
Assume we observe a sequence of multi-trial processes \(\{X_t^{(1)}, X_t^{(2)}, \ldots\}\) and aim to detect a regime change in the homological features. There exists a positive integer \(\eta\) such that:
\[
\left[S^{(1)}\right] = \ldots = \left[S^{(\eta)}\right] \neq \left[S^{(\eta+1)}\right] = \left[S^{(\eta+2)}\right] = \ldots
\]
\end{assumption}

\noindent In this context, we define the null and alternative hypotheses at trial \(r\) as follows:

\begin{eqnarray}
&& H_0: \left[S^{(l_0)}\right] = \ldots = \left[S^{(r+1)}\right],\\ %\quad l=l_0,\ldots,r,\\
&& H_1: \left[S^{(l_0)}\right] = \ldots = \left[S^{(r)}\right] \neq \left[S^{(r+1)}\right].% \quad l=l_0,\ldots,r,
\end{eqnarray}

To carry out this hypothesis testing in an online framework, we compute the spectral characteristics and summarize their topological features using persistence diagrams. We then assess the null hypothesis by evaluating the discrepancies between these spectral characteristics. This method utilizes the homological properties of spectral features to monitor changes in the equivalence classes of spectral characteristics over time. We employ two distinct approaches for hypothesis testing:

\begin{enumerate}
    \item \textbf{Past vs. Next Trial Comparison:} The first approach involves comparing the spectral characteristics from previous trials with those of the subsequent trial, assuming stationarity across trials. In this case, we set \( l_0 = 1 \), allowing us to compare the spectral characteristics of the past \( r \) trials with the next trial. % 

    \item \textbf{Present vs. Next Trial Comparison:} The second approach compares the spectral characteristics of the current trial with those of the following trial. This method is suitable when we anticipate smooth changes across trials. Here, we set \( l_0 = r \), which enables the comparison of only the most recent trial with the next.
\end{enumerate}

To evaluate a change in regime, we first need to summarize the spectral features and then define a measure of discrepancy between these summaries.

\begin{definition}[Spectral Features]
To quantify the discrepancy between features in successive trials, we define the following measures:
\begin{eqnarray}
    && D_r^{dgm} = \text{dist}\left(\text{dgm}\left(\text{Filt}(\hat{S}^{(r)})\right), \text{dgm}\left(\text{Filt}(\hat{S}^{(r+1)})\right)\right), \\
    && D_r =\text{dist}\left(\hat{S}^{(r)}, \hat{S}^{(r+1)}\right), 
\end{eqnarray}

\noi \(\text{dist}(\cdot,\cdot)\) represents a suitable metric for comparing the features, such as \(L_1\), \(L_2\), \(L_1\)-Wasserstein or \(L_2\)-Wasserstein distance when comparing persistence diagrams.

We define $\hat{S}^{(r)}$ in two possible manner:
\begin{itemize}
    \item
$$
\hat{S}^{(r)}=\frac{1}{r - l_0 + 1} \sum_{\ell = l_0}^{r} I_\ell,
$$
when the process is stationary within a trial and across trials, where $I_\ell$ is the periodogram computed on the process $X^{\ell}_t$ following Equation \ref{eq:periodogram}, where $1 \leq l_0 \leq r$.
    \item In other cases, $\hat{S}^{(r)} = I_r$ is the periodogram if weakly-stationary within the r-th trial or $\hat{S}^{(r)} = S_r$ is the spectrogram if the process is locally-stationary within the r-th trial.
\end{itemize}
\end{definition}

% To incorporate the spectral features of the next trial, we also express \(\widehat{T}_r\) in terms of the most recent trial:
% \[
% \widehat{T}_r = \frac{r}{r + 1} \widehat{T}_{r - 1} + \frac{1}{r + 1} T(\boldsymbol{X}^{(r)}).
% \]

The specific choice of distance metric will be discussed in the next section. These distance functions quantify the degree of dissimilarity between the observed class representations of the time series across trials. 

Under the null hypothesis that no regime change has occurred, the discrepancy measures \( D_r \) are expected to fluctuate randomly around a given mean value \( \mu_D \). When a regime change occurs, we expect a significant increase in the distance measure. To detect such changes, we apply the cumulative sum (CUSUM) method. The CUSUM statistic \( C_r \) at trial \( r \) is defined recursively as:
\begin{equation}
    C_r = \max\left(0, C_{r-1} + (D_r - \mu_D)\right),
    \label{eq:cusum_statistic}
\end{equation}
where \( \mu_D \) represents the expected mean of the distance measure under the assumption of no regime change, and \( C_0 = 0 \).

The CUSUM statistic accumulates deviations of the distance measure \( D_r \) from the expected mean \( \mu_D \), resetting to zero whenever the cumulative sum becomes negative. This allows us to detect sustained deviations from the mean, which are indicative of a regime change. A regime change is declared when the CUSUM statistic \( C_r \) exceeds a predetermined threshold \( h \). Formally, a regime change at trial \( r \) is detected if:

\begin{equation}
    C_r > h.
    \label{eq:cusum_detection}
\end{equation}
The threshold \( h \) is chosen based on the desired trade-off between sensitivity and false positive rates. A smaller threshold \( h \) increases the sensitivity to smaller shifts but may result in more false positives, whereas a larger \( h \) detects only larger deviations, reducing false alarms but potentially missing smaller regime changes.

% {\color{red} choice of the threshold ongoing work}

% \begin{remark}
% In contrast to the Slowly Evolving Locally Stationary Process studied in \cite{DYNAMIC_BRAIN_PROCESSES}, our model allows also for abrupt changes from one trial $r$ to another. 
% \end{remark}

% CP problem in standard case:

%Remark for us, index both process Z and transfer function on $r$ , the spectral characteristics might be different and the processes generating at each replicated independent, for simplicity. We allow for change at each trials (thinking that the brain regime might be the on a translated / dilated time scale . we specifically wanna adapt to that define set of spectral densities, time-varying spectral density  / group action 

% We present our approach to detecting regime changes under three distinct scenarios, each motivated by practical circumstances encountered in time series analysis. The scenarios are as follows:

%put different scenario in simulation section  . Check in scenario 2 , the test statistic is based on a distance between periodograms (they have to be smoothed since the periodogram is not consistent) , no problem when it concerns spectrograms
% Yes, indeed. I have smoothed the spectrograms ... I will make this more explicit.

\section{Numerical experiments}
\label{sec:simulations}

This section presents multiple examples to illustrate the effectiveness of our approach to regime change detection across the different scenarios discussed previously. We will provide a brief overview of how the data was generated and analyzed, followed by a comparative analysis of the metrics we propose.

\subsection{Stationary Time Series within Trials and Across Trials} 

In this scenario, we assume that the time series is stationary both within individual trials and across multiple trials. This means that there are no significant changes in the statistical properties of the time series until a regime change occurs. Such a scenario is often observed in controlled experimental settings where conditions remain constant over time, and any deviation indicates a significant shift in the underlying process. We explore two distinct examples of regime change, each characterized by different underlying processes.

\textbf{Example 1: From Low- to High-Frequency Regime Change} \\
In the first example, we generated a dataset comprising 200 trials of AR(2) processes. The initial 100 trials exhibit a predominant low-frequency component with peak at 10 Hz, while the subsequent 100 trials introduce a regime change characterized by the inclusion of a high-frequency component at 40 Hz. Each trial consists of 2 seconds of data with 100 Hz sampling rate similar to examples in Figure \ref{fig:AR2_PSD}.

\textbf{Example 2: From Low- to Mixed-Frequency Regime Change} \\
The second example follows a similar structure, with the first 100 trials also representing a low-frequency AR(2) process. However, in this scenario, the following 100 trials incorporate mixed frequencies, both the low and high frequencies. This setup aims to assess and contrast the robustness of our proposed metrics in detecting regime changes.

\begin{figure}[H]
    \centering
    \includegraphics[width=\textwidth]{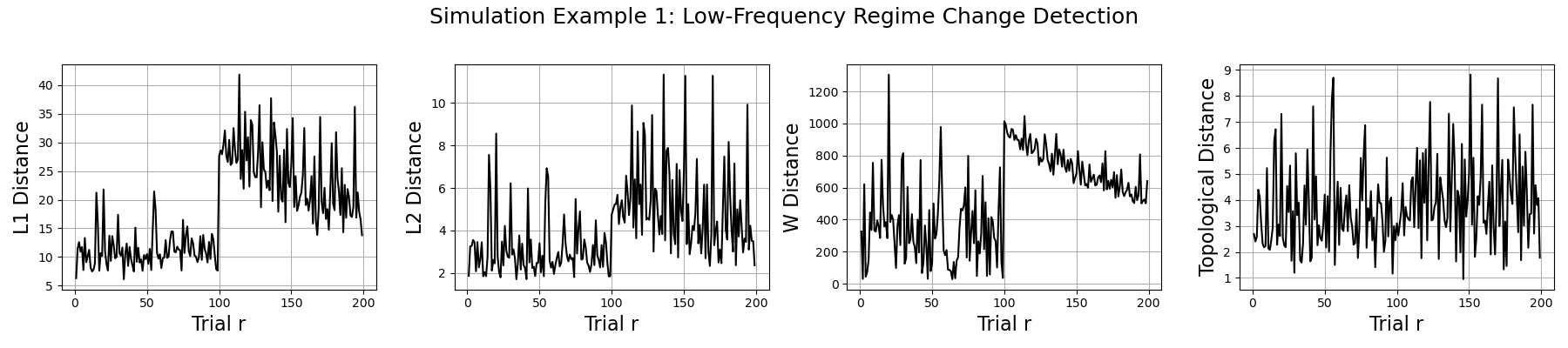}
    \caption{Distance evolution between consecutive trials in Scenario 1, analyzed using different metrics. From left to right: L1, L2, L1-Wasserstein distance between periodograms, and L2-Wasserstein distance between consecutive \( H_0 \) diagrams.}
    \label{fig:simulation_example_1}
\end{figure}

\begin{figure}[H]
    \centering
    \includegraphics[width=\textwidth]{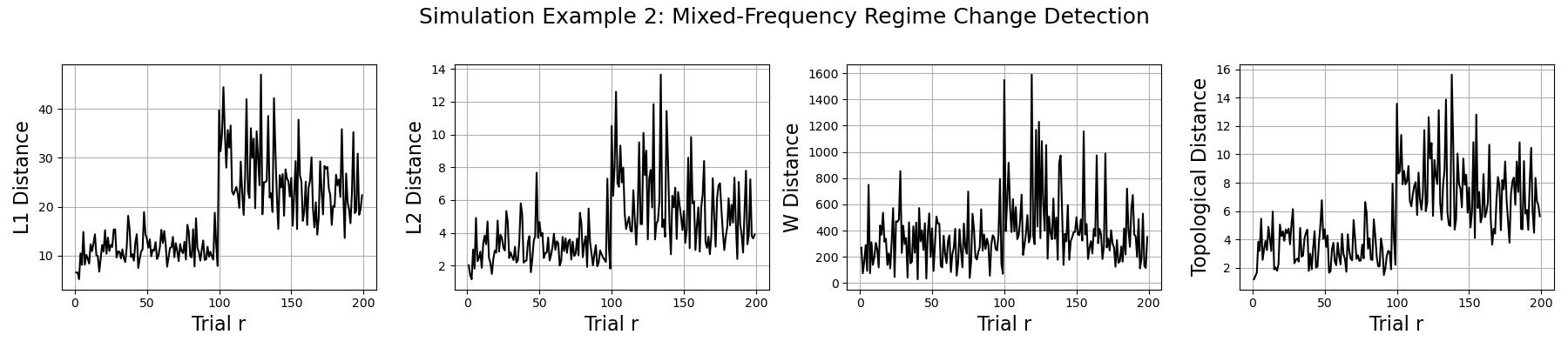}
    \caption{Distance evolution between consecutive trials in Scenario 1, analyzed using different metrics. From left to right: L1, L2, L1-Wasserstein distance between periodograms, and L2-Wasserstein distance between consecutive \( H_0 \) diagrams.}
    \label{fig:simulation_example_2}
\end{figure}

In both examples, assuming stationarity across trials, we combine information from past trials to obtain a more precise estimate of the cumulative behavior. We computed distances between the estimated PSDs up to trial \( r \) and at \( r+1 \) using various metrics: L1, L2, Wasserstein (W), and Topological (T). The results of our analysis are illustrated in Figures \ref{fig:simulation_example_1} and \ref{fig:simulation_example_2}.

The analysis reveals that the L1 metric consistently captures regime changes across both examples. The L2 metric also detects changes; however, its effectiveness is less pronounced than that of L1, particularly in Example 1. Notably, the L2 metric exhibits improved performance in Example 2 compared to Example 1. The Wasserstein distance performs strongly in Example 1 but shows diminished effectiveness in Example 2, primarily because it measures the Earth Mover's Distance between the two spectra. Since the first half of the spectrum remains unchanged, this results in a relatively smaller distance. Conversely, the Topological distance performs well in Example 2, as regime changes introduce two peaks, making the spectrum topologically different across trials. In contrast, both low- and high-frequency AR(2) processes in Example 1 yield similar PSD shapes, leading to comparable persistence diagrams which leads to stable topological distance.

\subsection{Stationary Time Series within Trials and Non-Stationary Across Trials} 

Here, we assume that while the time series remains stationary within each trial, its statistical properties can vary across different trials. These variations across trials are assumed to be smooth and gradual until a regime change occurs. This scenario is applicable in experimental setups where conditions change systematically between sessions, such as gradual learning or adaptation processes, but remain consistent within a single session.

\textbf{Example 3: Evolving spectrum.} \\

\begin{figure}[H]
    \centering
    \includegraphics[width=\linewidth]{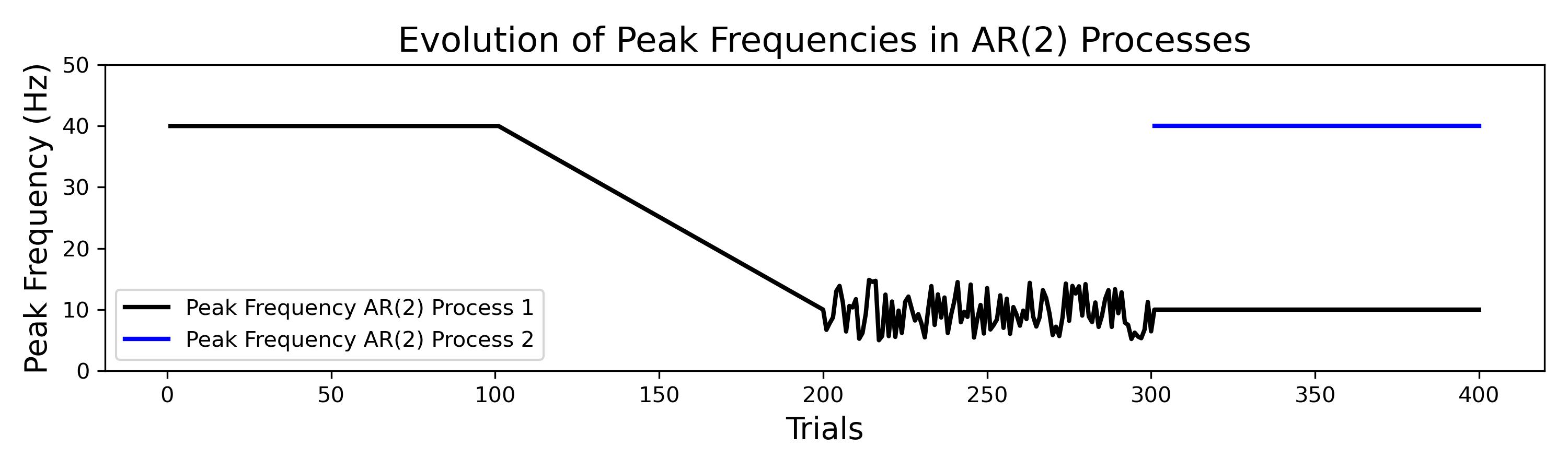}
    \caption{Illustration of the spectral properties of AR(2) processes across trials.}
    \label{fig:ar2_peak_frequency_evolution}
\end{figure}

The goal of this example is to detect changes in the spectral properties of time series as they evolve through trials. We consider a time series comprising 400 trials, each lasting 5 seconds and sampled at a rate of 100 Hz. In the initial 100 trials, we observe an AR(2) process with a fixed peak frequency of 40 Hz. During the subsequent 100 trials, this peak frequency linearly transitions from 40 Hz down to 10 Hz. The next segment consists of 100 trials where the AR(2) process exhibits a random peak frequency fluctuating between 5 Hz and 15 Hz. Finally, in the last 100 trials, a regime change occurs and the process becomes a mixture of two AR(2) processes, characterized by peak frequencies at 10 Hz and 40 Hz. Figure \ref{fig:ar2_peak_frequency_evolution} illustrates the evolution of these peak frequencies across the trials, providing a visual representation of the evolving spectral properties of the observed time series.

The analysis of the results highlights significant variations in the performance of different metrics for detecting regime changes. Notably, the Wasserstein metric is the least effective in identifying substantial transitions. In contrast, both the L1 and L2 metrics successfully avoid signaling changes between the first and second segments of trials. Interestingly, the topological metric excels by accurately capturing the regime change after trial 300, which coincides with the introduction of a mixture of two AR(2) processes. However, the L1 and L2 metrics exhibit an intriguing behavior by falsely indicating a regime change around trial 200. This false positive can be attributed to the randomness introduced by the rapidly varying peak frequencies of the AR(2) process during that segment. In contrast, the topological metric demonstrates robustness to frequency shifts, successfully detecting the presence of the additional AR(2) process. This detection occurs as the introduction of another peak alters the sublevel set filtration of the spectrum.

\begin{figure}[H]
    \centering
    \includegraphics[width=\linewidth]{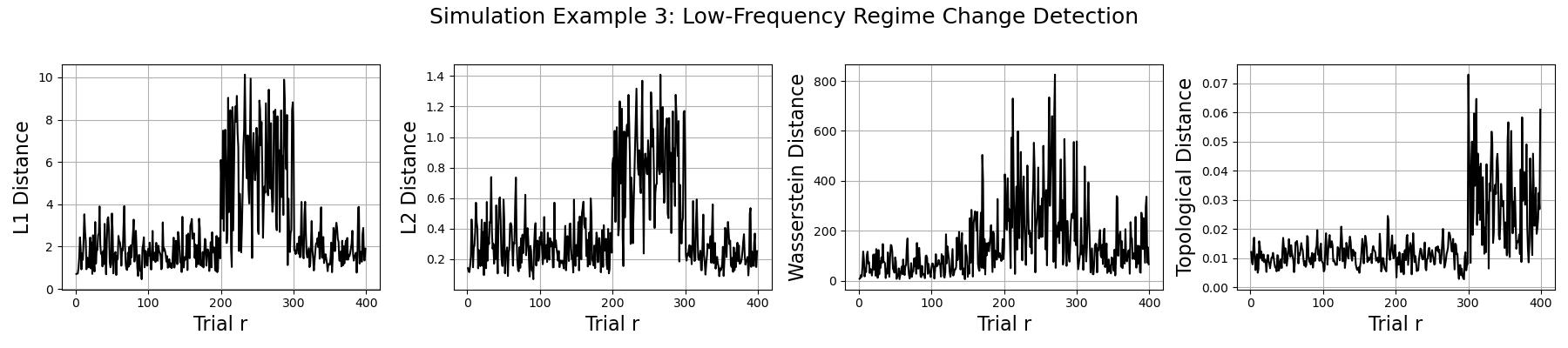}
    \caption{Distance evolution between consecutive trials in Scenario 2, analyzed using different metrics. From left to right: L1, L2, L1-Wasserstein distance between periodograms, and L2-Wasserstein distance between consecutive \( H_0 \) diagrams.}
    \label{fig:simulation_example_3}
\end{figure}

\subsection{Locally Stationary Time Series within Trials and Non-Stationary Across Trials}

In this scenario, we consider the most complex case where the time series exhibits local stationarity within trials, meaning that its statistical properties change slowly over time within each trial. Additionally, there can be non-stationary behavior across trials. These changes, both within and across trials, are gradual until a significant regime change disrupts the established patterns. This scenario models real-world processes where underlying conditions evolve over time, such as fatigue in cognitive experiments or gradual degradation in mechanical systems.

\textbf{Example 4: Evolving Spectrograms} \\

In this example, we analyze how the spectral characteristics of an observed time series, $X_r(t)$, evolve across trials ($r$) and within individual trials over time ($t$). Our goal is to detect regime changes by studying these evolving spectral properties. The dataset consists of 200 trials, each lasting 15 seconds and sampled at 100 Hz. 

\begin{figure}[H]
    \centering
    \includegraphics[width=\textwidth]{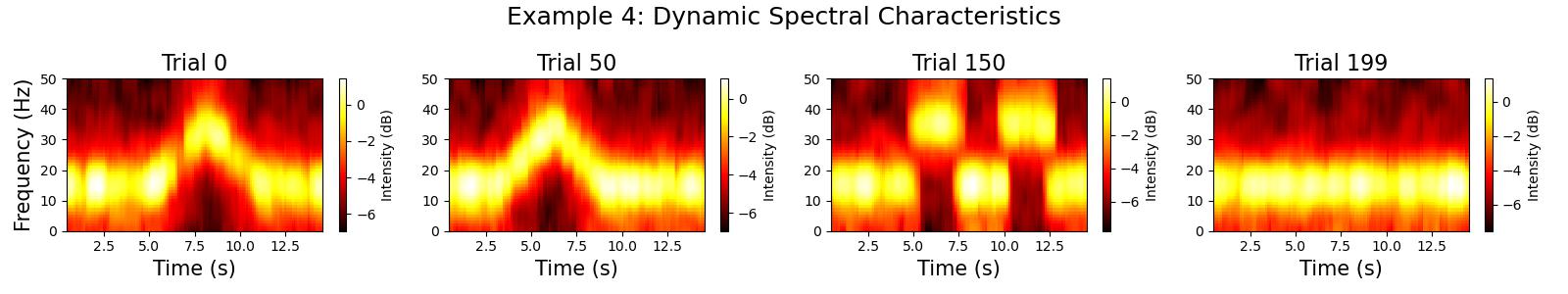}
    \caption{Evolving time-varying spectrum across trials in Example 4.}
    \label{fig:evolving_spectrogram_scenario_4}
\end{figure}

In the first 100 trials, the time series follows a time-varying AR(2) process. The peak frequency increases smoothly from 15 Hz to 35 Hz and then returns to 15 Hz. This change occurs at a randomly chosen point between 2 and 9 seconds into the trial, and the entire rise and fall take approximately 6 seconds. In the subsequent 100 trials, the AR(2) process alternates between two scenarios: one with a fixed peak frequency of 15 Hz, and the other where the peak frequency abruptly transitions between 15 Hz and 35 Hz multiple times within a trial. Figure \ref{fig:evolving_spectrogram_scenario_4} illustrates the evolution of the spectrograms across trials.

\begin{figure}[H]
    \centering
    \includegraphics[width=\textwidth]{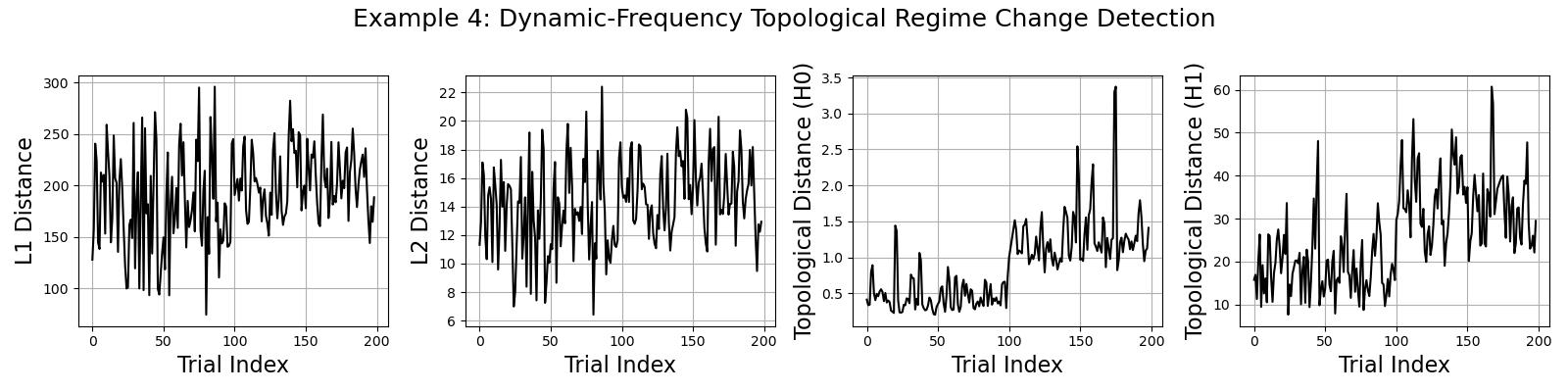}
    \caption{Distance evolution between consecutive trials in Scenario 3, analyzed using different metrics. From left to right: L1, L2, L2-Wasserstein distance between consecutive \( H_0 \) and \( H_1 \) diagrams.}
    \label{fig:simulation_example_4}
\end{figure}

We also compare consecutive trials using multiple distance metrics to detect changes. Figure \ref{fig:simulation_example_4} shows the results for various metrics. While the L1 and L2 metrics fail to capture the regime change effectively, the topological Wasserstein metric between persistence diagrams highlights a notable increase in topological variability as we approach the second half of the trials. 

%In the first 100 trials, the spectral topological variations are relatively small, since the differences largely consist of horizontal translations of the peak frequency (e.g., varying onset times of the event). Consequently, the topological metric detects only minor changes. However, in the second half of the trials, the introduction and disappearance of distinct peaks alter the spectral topology significantly, resulting in greater topological variability both in \( H_0\) (disconnected components) and \( H_1\) (holes) Homology.

In the first 100 trials, the spectral variations primarily involve horizontal translations of the peak frequency (e.g., varying onset times of the event). Consequently, the topological distance randomly fluctuates but do not reveal any major change. However, in the second half of the trials, the introduction, and disappearance of distinct peaks significantly alter the spectral topology. This results in greater topological variability both in \( H_0\) (disconnected components) and \( H_1\) (holes) Homology.

\section{Analysis of Accelerometry Vibration Signals}
\label{sec:application}

\begin{figure}[H]
    \centering    \includegraphics[width=0.7\textwidth]{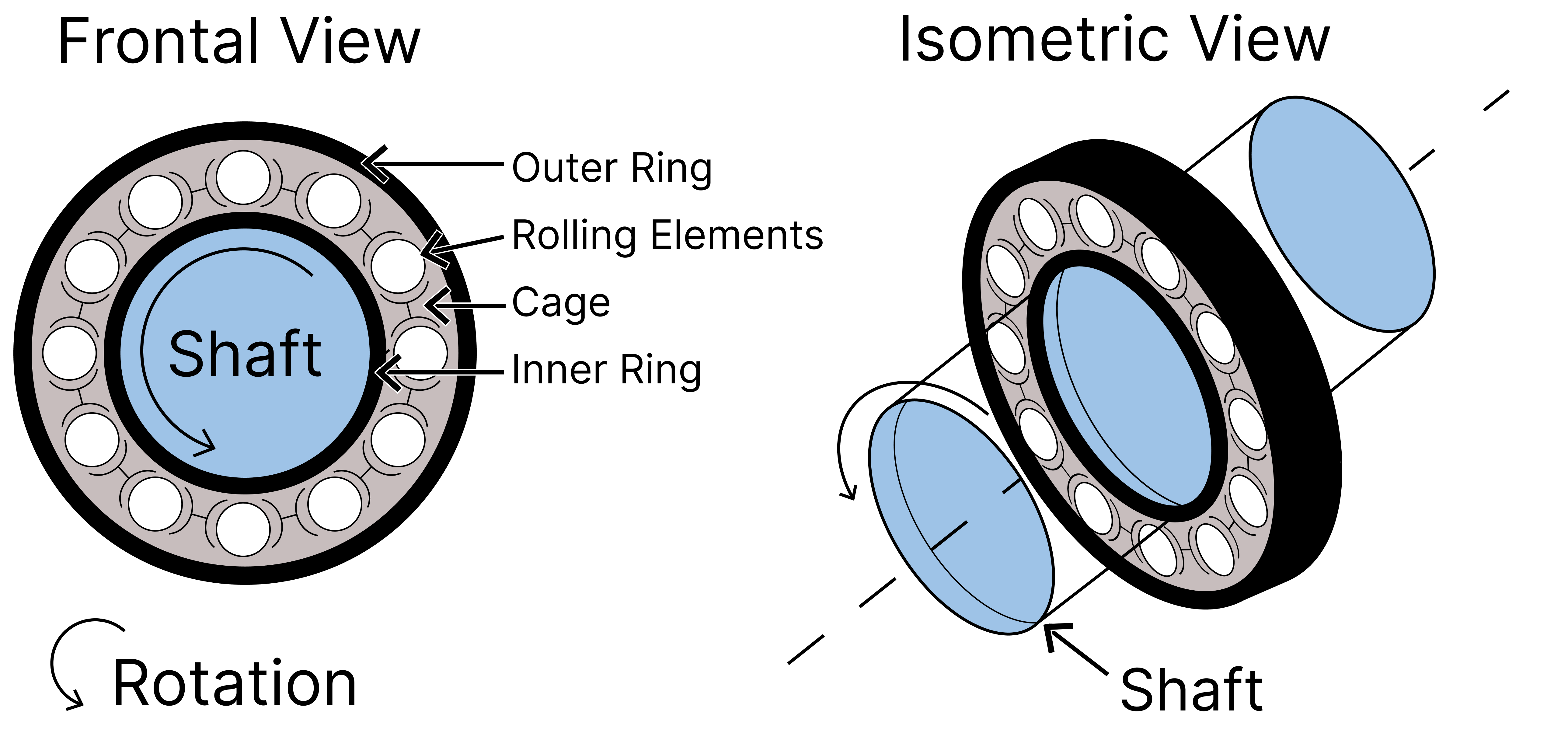}
    \caption{Illustration of a bearing and its function in supporting a rotating shaft.}
    \label{fig:bearing_illustration}
\end{figure}

In the field of predictive maintenance, monitoring the behavior of critical machine components is essential for detecting early signs of fatigue. Vibration data, in particular, is commonly collected from sensitive components that pose a risk of mechanical failure. Bearings play a crucial role in facilitating the rotation of objects by supporting the shaft, as illustrated in Figure \ref{fig:bearing_illustration}. These components are vital to the operation of most rotating machinery and are designed to function for a specified number of revolutions, typically measured in millions of revolutions. However, cumulative shocks and excessive loads can lead to gradual wear and eventual failure. The failure of a bearing often results in costly breakdowns and operational downtime.
To mitigate these risks, monitoring techniques such as accelerometers are employed to measure vibrations. These tools enable engineers to detect abnormal behavior in the machinery, potentially identifying early signs of tear before catastrophic failures occur. This proactive approach allows for the timely replacement of at-risk or damaged components, thereby enhancing machine reliability and reducing operational and maintenance costs.

\begin{figure}[H]
    \centering
\includegraphics[width=0.7\textwidth]{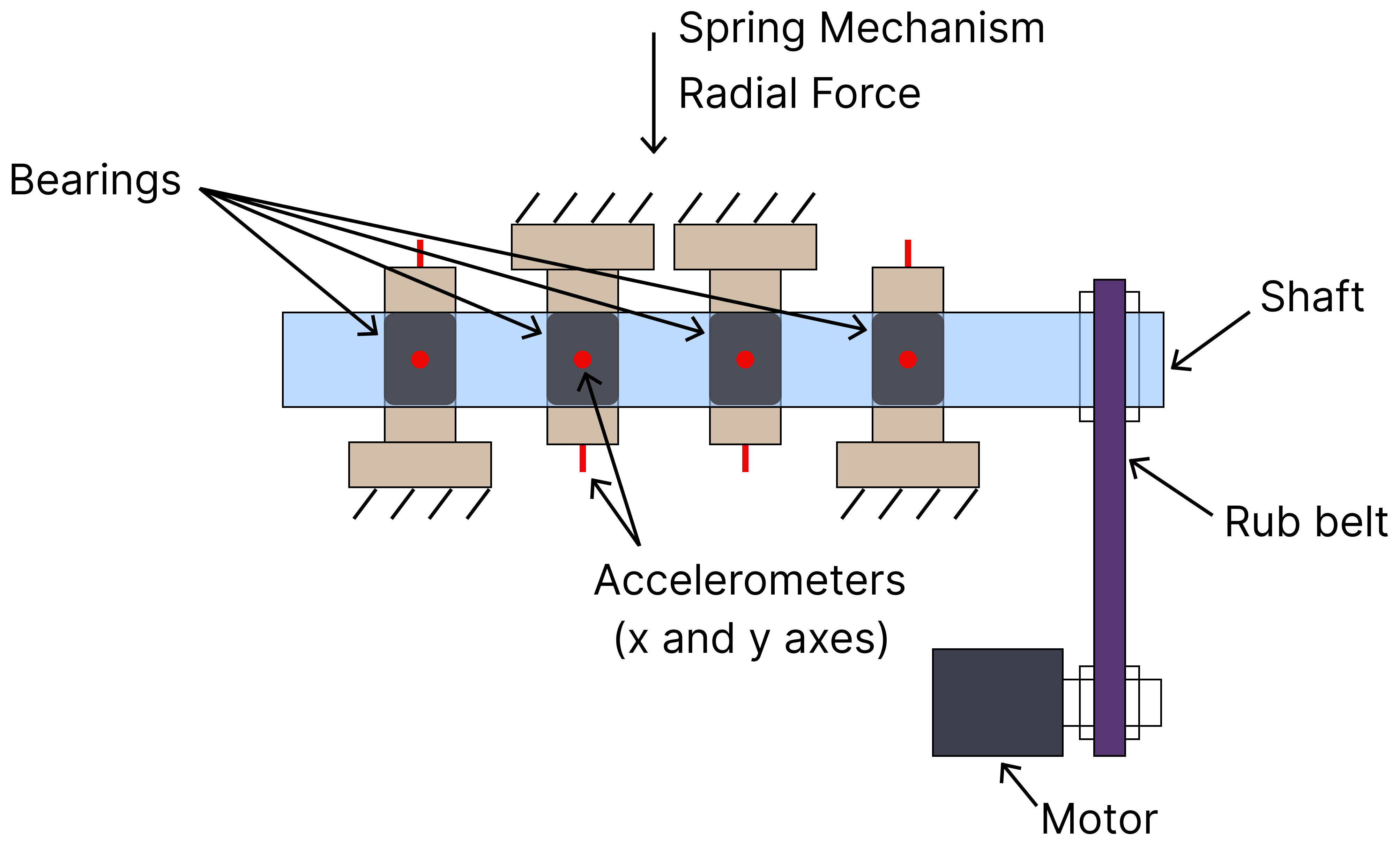}
    \caption{Schematic representation of the bearing failure experiment setup. The four bearings are mounted on a shaft subjected to a radial load.}
    \label{fig:bearing_failure_experiment_illustration}
\end{figure}

In this study, we analyze the NASA IMS bearing data, which consists of a test-to-failure experiment involving four bearings mounted on a shaft driven by an AC motor operating at a constant speed of 2000 RPM \citep{BEARING_DATA_SET}. A radial load of 6,000 pounds (2.72 ton) is applied to the shaft and bearings using a spring mechanism, as shown in Figure \ref{fig:bearing_failure_experiment_illustration}. The dataset captures vibration signal snapshots recorded at specific intervals (every 5 or 10 minutes) over the course of the experiment, with each trial lasting one second. Each file contains 20,480 measurements sampled at 20 kHz.

\begin{figure}[H]
    \centering
    \includegraphics[width=\textwidth]{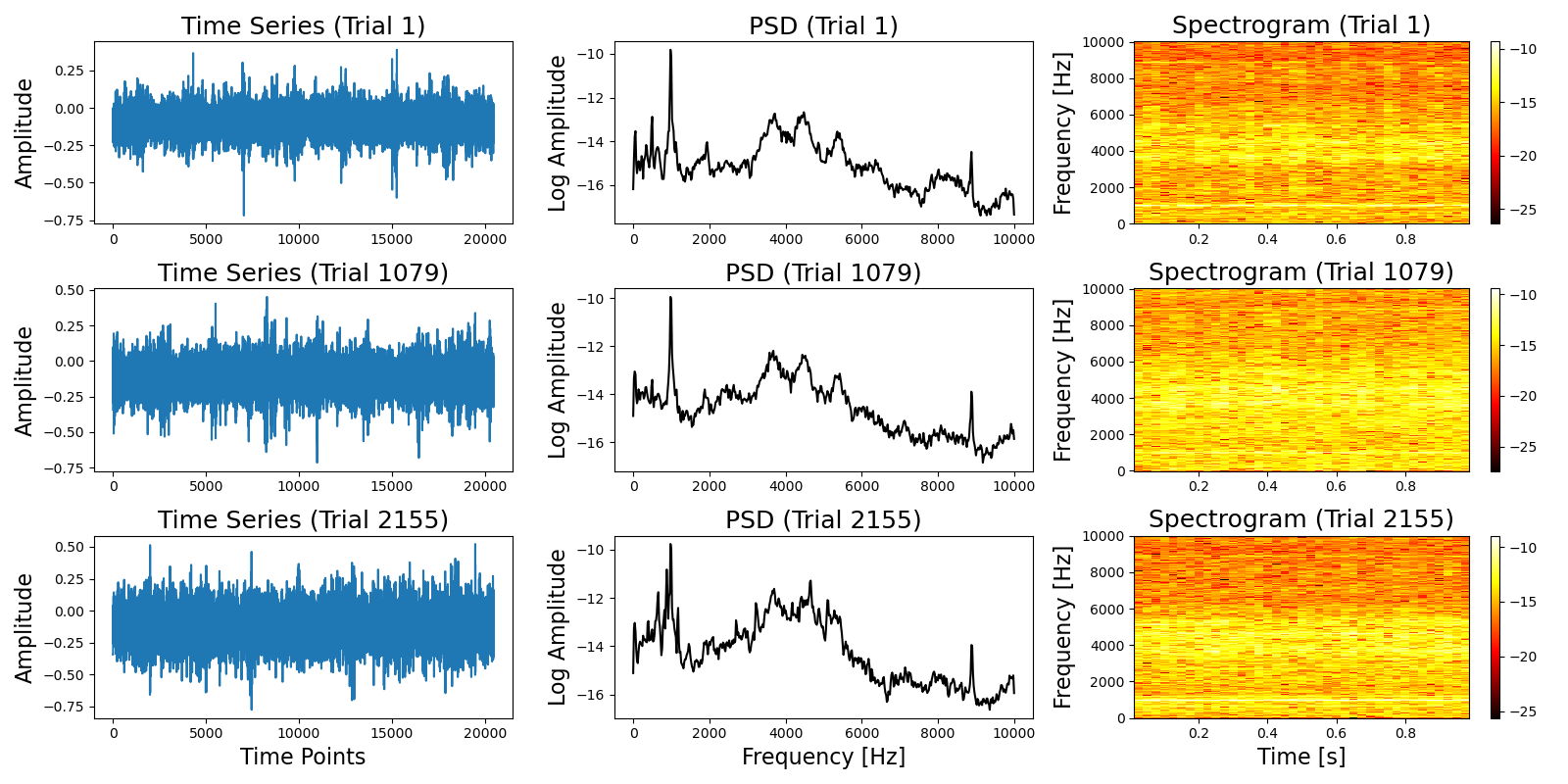}
    \caption{Visualization of three trials from the beginning, middle, and end of the test-to-failure's first experiment of bearing number one. Left: the observed accelerometer vibrational measures time series. Middle: the estimated spectra. Right: spectrogram.}
    \label{fig:bearing_data_1}
\end{figure}

\begin{figure}[H]
    \centering
    \includegraphics[width=\textwidth]{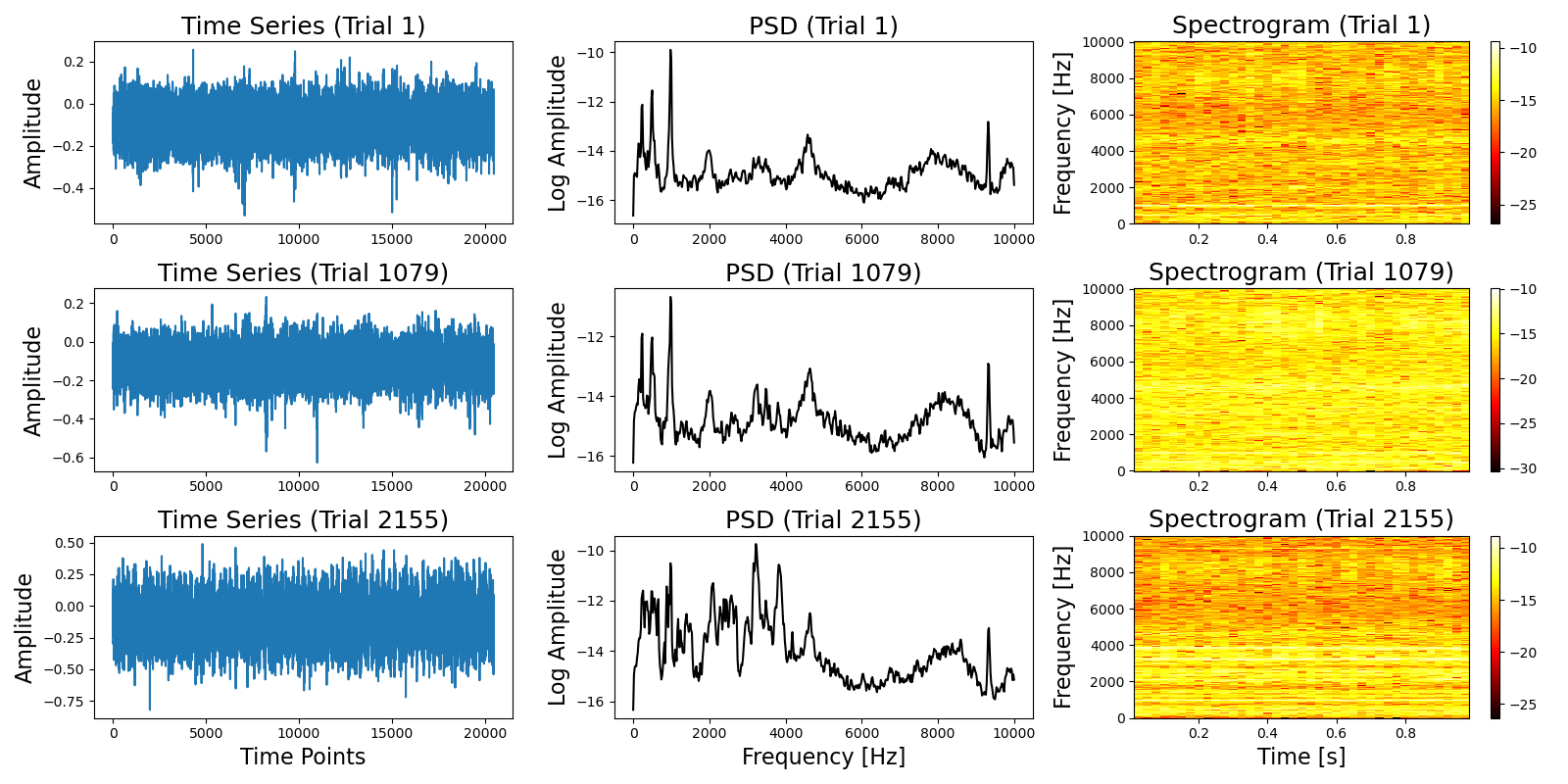}
    \caption{Visualization of three trials from the beginning, middle, and end of the test-to-failure's second experiment and of bearing number one. Left: observed accelerometer vibrational measures time series. Middle: the estimated spectrum. Right: spectrogram.}
    \label{fig:bearing_data_2}
\end{figure}

Using our proposed approach, we analyzed the data from two test-to-failure experiments. Each recorded vibration signal constitutes a one-dimensional time series, with each trial consisting of \( T = 20480 \) observations, as can be seen in Figures \ref{fig:bearing_data_1} and \ref{fig:bearing_data_2}. The first experiment encompasses \( R = 2156 \) trials, while the second experiment includes \( R = 984 \) trials.


\begin{figure}[H]
    \centering
    \includegraphics[width=\textwidth]{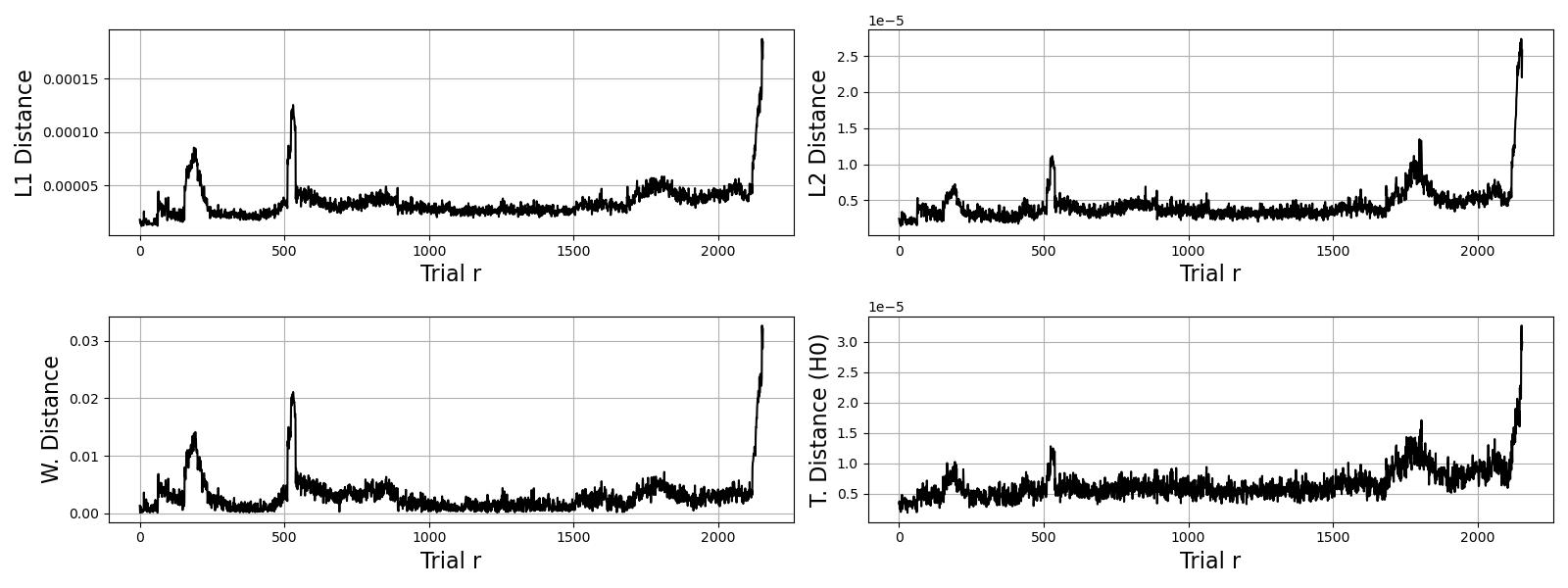}
    \caption{Distance evolution for bearing one under scenario 1. The figure illustrates the metrics L1, L2, Wasserstein, and topological distance between \( H_0 \) across the trial.}
    \label{fig:scenario1_bearing1}
\end{figure}

\begin{figure}[H]
    \centering
    \includegraphics[width=\textwidth]{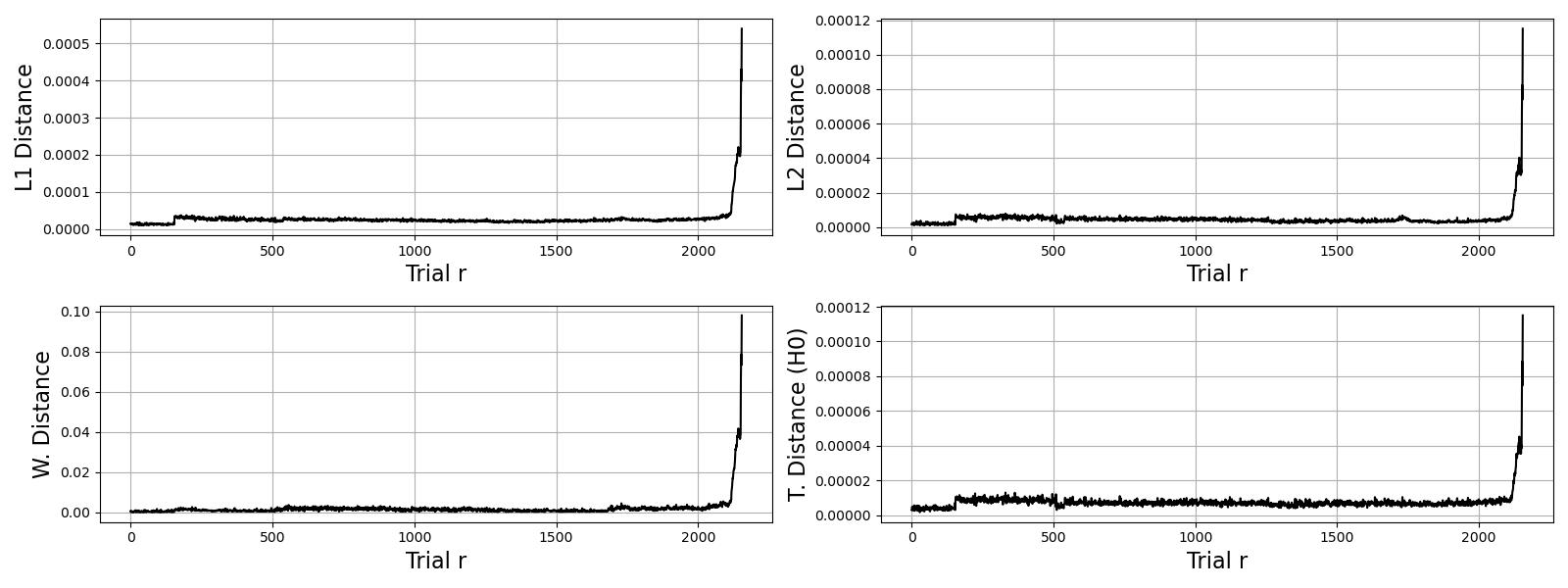}
    \caption{Distance evolution for bearing four under scenario 1. This figure presents the L1, L2, Wasserstein, and topological distance between \( H_0 \) across the trial.}
    \label{fig:scenario1_bearing4}
\end{figure}

\begin{figure}[H]
    \centering
    \includegraphics[width=\textwidth]{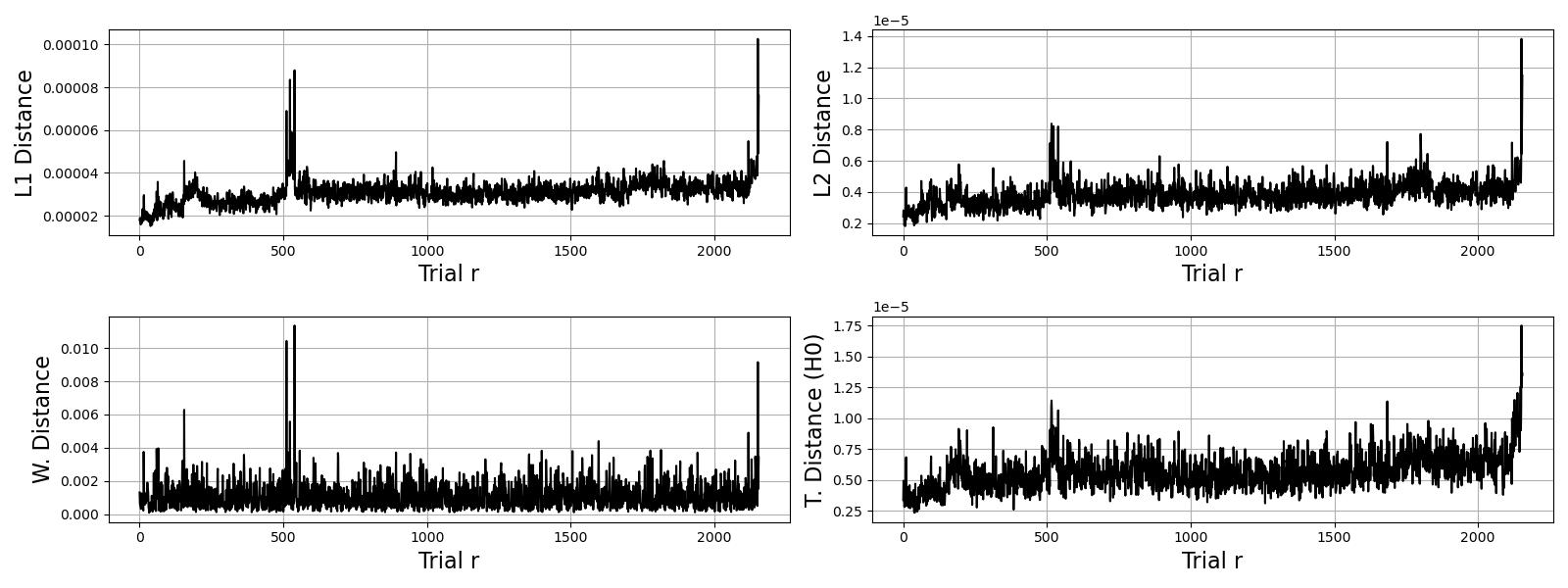}
    \caption{Distance evolution for bearing one under scenario 2. The figure depicts the changes in L1, L2, Wasserstein, and topological distance between \( H_0 \) across the trial.}
    \label{fig:scenario2_bearing1}
\end{figure}

\begin{figure}[H]
    \centering
    \includegraphics[width=\textwidth]{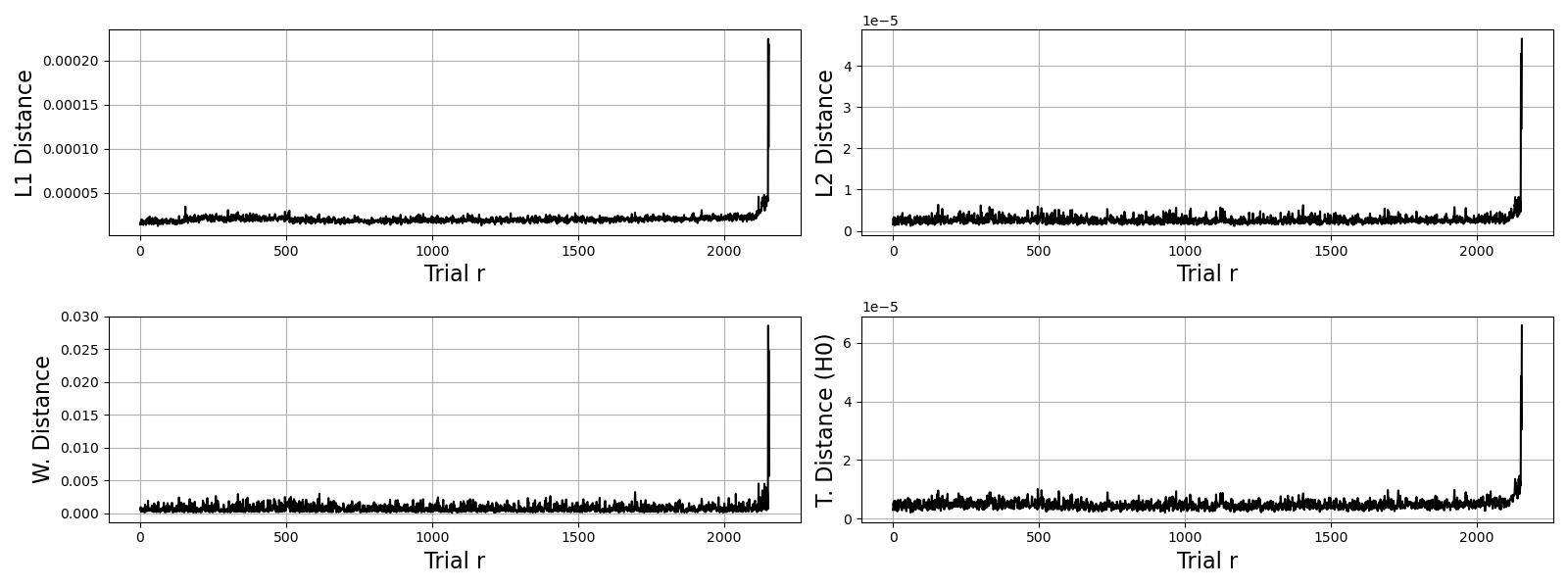}
    \caption{Distance evolution for bearing four under scenario 2. The figure illustrates the variations in L1, L2, Wasserstein, and topological distance between \( H_0 \) across the trial.}
    \label{fig:scenario2_bearing4}
\end{figure}

The change of regime analysis of the previous data from the first experiment of bearing one (see Figures \ref{fig:scenario1_bearing1}, \ref{fig:scenario2_bearing1}) and four (see Figures \ref{fig:scenario1_bearing4}, \ref{fig:scenario2_bearing4}) leads to the following results. We can see that all the proposed metrics are able to detect the dynamics of the data during the experiment. In bearing one, we observe increases in the distance metrics quite early in the trial; this is probably due to changes in the load during the experiment or some experimental error. Towards the end of the experiment, we see an increase, but it is not extreme compared to the beginning. 

However, in the fourth bearing, we see a different behavior as the measures are quite stable. However, towards the end, there is a drastic increase that indicates a change of regime. This aligns well with the experimental results, as we know that at the end of the test-to-failure experiment, an inner race defect occurred in bearing 3 and a roller element defect in bearing 4, but not in bearings 1 and 2.
Furthermore, we observe significant differences in the results between scenarios 1 and 2. In scenario 2, the distance estimates exhibit more abrupt changes, with localized peaks that indicate precise timing of changes during the experiment. In contrast, scenario 1 shows more gradual trends and a smoother evolution of distances. This discrepancy arises from the underlying assumptions of each scenario.

Specifically, scenario 1 assumes stationarity within trials and across trials, meaning that if a change in spectral properties occurs, it affects the average over a longer period. Conversely, scenario 2 focuses on comparing consecutive trials, thereby ignoring any influences from previous trials. This approach allows it to capture more immediate changes but at the expense of potential context provided by the history of prior trials which leads to increased "fuzziness" in the distance estimates.

\begin{figure}[H]
    \centering
    \includegraphics[width=\textwidth]{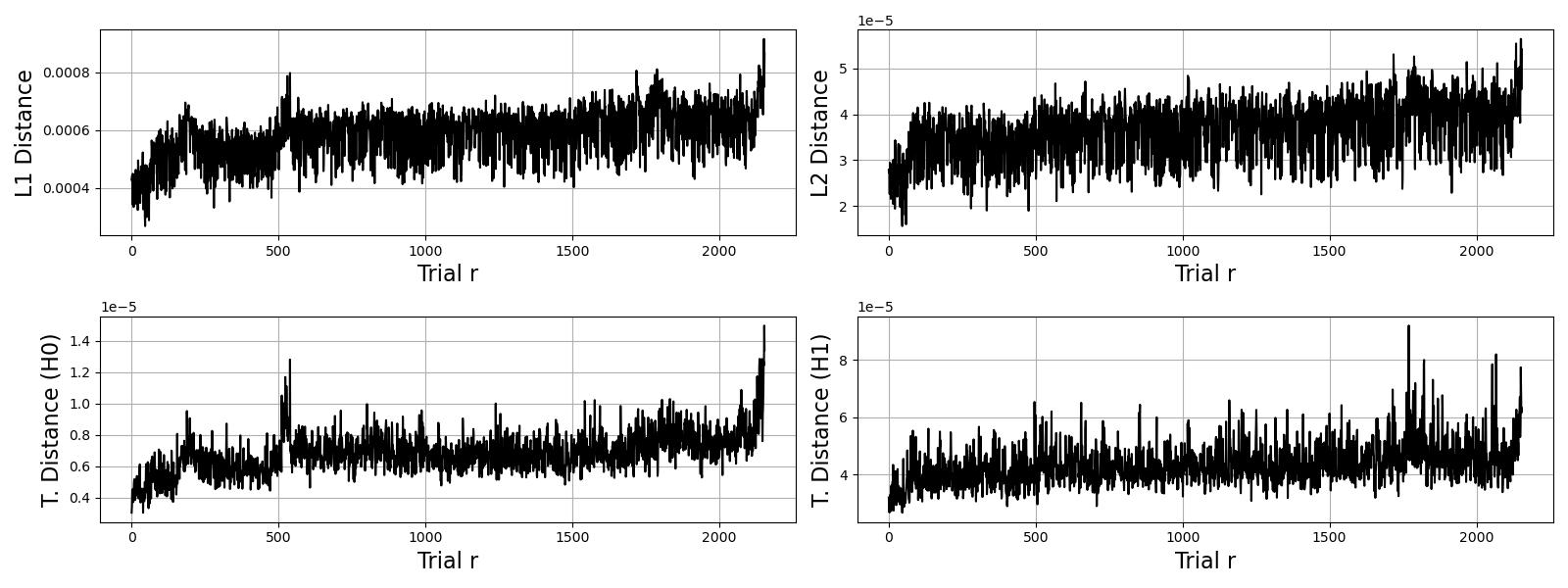}
    \caption{Distance evolution for varying metrics through trials of bearing 1 under scenario 3.}
    \label{fig:scenario3_bearing1}
\end{figure}

\begin{figure}[H]
    \centering
    \includegraphics[width=\textwidth]{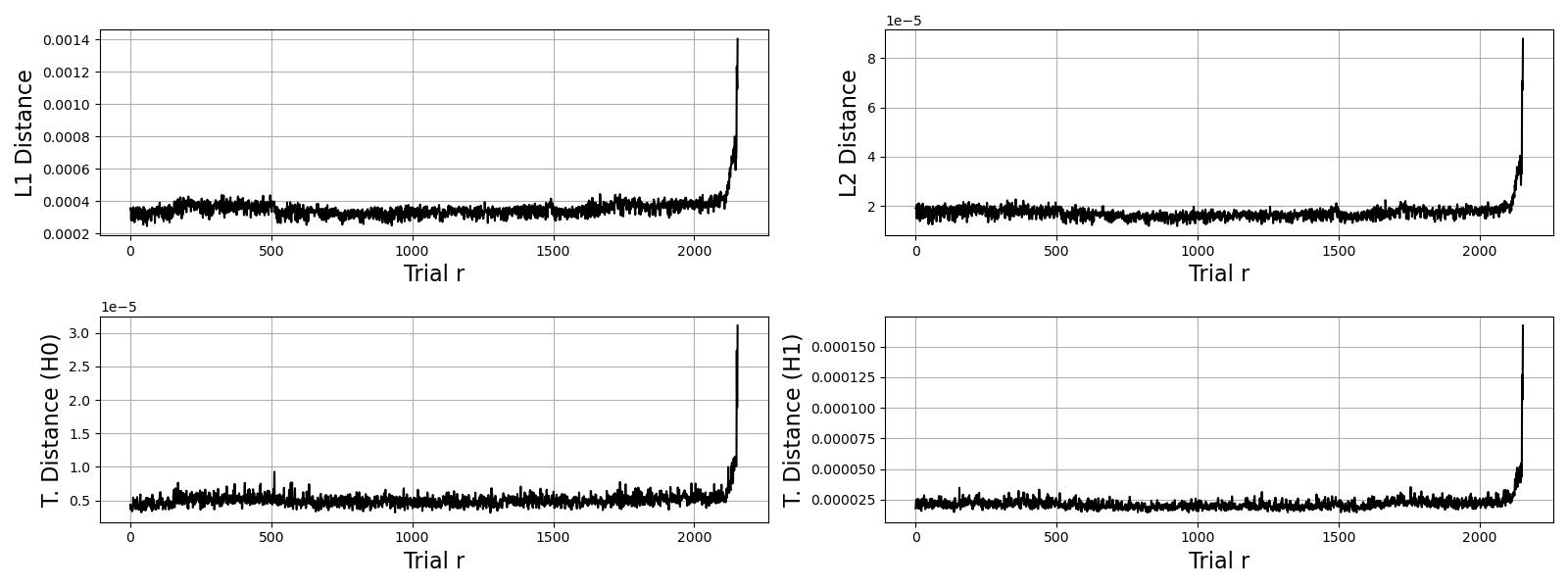}
    \caption{Distance evolution for varying metrics through trials of bearing 4 under scenario 3.}
    \label{fig:scenario3_bearing4}
\end{figure}

In the third scenario, since the time series data is assumed to be non-stationary both within and across trials, the spectrogram is employed to estimate the spectral characteristics of the time series. In Figures \ref{fig:scenario3_bearing1} and \ref{fig:scenario3_bearing4}, we observe the evolution of the distances based on the various metrics employed. A notable distinction in behavior is evident between bearings 1 and 4. For bearing 1, the distance fluctuates consistently throughout the trials, suggesting transient changes in the spectral properties. However, in bearing 4, we see a drastic increase towards the end of the experiment, signaling a regime change that coincides with the bearing's failure. This abrupt shift is captured by all metrics, underlining their effectiveness in detecting the regime shift. 

\section{Conclusion}
\label{sec:conclusion}

This paper introduced a novel and flexible framework for online detection of regime changes in multi-trial, non-stationary time series data. Unlike traditional methods that focus solely on identifying change points, our approach is designed to handle more complex scenarios where time series may be non-stationary both within and across trials. By leveraging a combination of metrics—including L1 and L2 norms, the Wasserstein distance, and topological features of time-frequency characteristics—our method can ignore some irrelevant changes and detect key topological changes in the spectral properties.

Our simulations demonstrated the robustness and adaptability of this framework in a variety of scenarios. Additionally, applying our method to the NASA Bearing dataset revealed insightful results: the detected regime changes aligned with potential signs of mechanical failures or early indications of fatigue. This underscores the potential of our approach for predictive maintenance and other applications where topological changes in underlying processes matter more than isolated change points.

\newpage
\bibliography{references}

\begin{thebibliography}{28}
\providecommand{\natexlab}[1]{#1}
\providecommand{\url}[1]{\texttt{#1}}
\expandafter\ifx\csname urlstyle\endcsname\relax
  \providecommand{\doi}[1]{doi: #1}\else
  \providecommand{\doi}{doi: \begingroup \urlstyle{rm}\Url}\fi

\bibitem[Aston and Kirch(2012)]{ASTON2012204}
John~A.D. Aston and Claudia Kirch.
\newblock Detecting and estimating changes in dependent functional data.
\newblock \emph{Journal of Multivariate Analysis}, 109:\penalty0 204--220, 2012.
\newblock ISSN 0047-259X.
\newblock \doi{https://doi.org/10.1016/j.jmva.2012.03.006}.
\newblock URL \url{https://www.sciencedirect.com/science/article/pii/S0047259X12000759}.

\bibitem[Aue et~al.(2009)Aue, Gabrys, Horváth, and Kokoszka]{AUE20092254}
Alexander Aue, Robertas Gabrys, Lajos Horváth, and Piotr Kokoszka.
\newblock Estimation of a change-point in the mean function of functional data.
\newblock \emph{Journal of Multivariate Analysis}, 100\penalty0 (10):\penalty0 2254--2269, 2009.
\newblock ISSN 0047-259X.
\newblock \doi{https://doi.org/10.1016/j.jmva.2009.04.001}.
\newblock URL \url{https://www.sciencedirect.com/science/article/pii/S0047259X09000827}.

\bibitem[Avanesov and Buzun(2018)]{Avanesov2018}
Valeriy Avanesov and Nazar Buzun.
\newblock {Change-point detection in high-dimensional covariance structure}.
\newblock \emph{Electronic Journal of Statistics}, 12\penalty0 (2):\penalty0 3254 -- 3294, 2018.
\newblock \doi{10.1214/18-EJS1484}.
\newblock URL \url{https://doi.org/10.1214/18-EJS1484}.

\bibitem[Barbarossa and Sardellitti(2020{\natexlab{a}})]{barbarossaTopologicalSignalProcessing2020}
Sergio Barbarossa and Stefania Sardellitti.
\newblock Topological {{Signal Processing Over Simplicial Complexes}}.
\newblock \emph{IEEE Transactions on Signal Processing}, 68:\penalty0 2992--3007, 2020{\natexlab{a}}.
\newblock ISSN 1941-0476.
\newblock \doi{10.1109/TSP.2020.2981920}.

\bibitem[Barbarossa and Sardellitti(2020{\natexlab{b}})]{barbarossaTopologicalSignalProcessing2020a}
Sergio Barbarossa and Stefania Sardellitti.
\newblock Topological {{Signal Processing}}: {{Making Sense}} of {{Data Building}} on {{Multiway Relations}}.
\newblock \emph{IEEE Signal Processing Magazine}, 37\penalty0 (6):\penalty0 174--183, November 2020{\natexlab{b}}.
\newblock ISSN 1053-5888, 1558-0792.
\newblock \doi{10.1109/MSP.2020.3014067}.

\bibitem[Bonafos et~al.(2023)Bonafos, Freyermuth, Pudlo, Tronçon, and Rey]{Guillem2}
Guillem Bonafos, Jean-Marc Freyermuth, Pierre Pudlo, Samuel Tronçon, and Arnaud Rey.
\newblock Topological data analysis of human vowels: Persistent homologies across representation spaces.
\newblock Technical report, 2023.

\bibitem[Casini and Perron(2024{\natexlab{a}})]{CASINI2024105811}
Alessandro Casini and Pierre Perron.
\newblock Change-point analysis of time series with evolutionary spectra.
\newblock \emph{Journal of Econometrics}, 242\penalty0 (2):\penalty0 105811, 2024{\natexlab{a}}.
\newblock ISSN 0304-4076.
\newblock \doi{https://doi.org/10.1016/j.jeconom.2024.105811}.

\bibitem[Casini and Perron(2024{\natexlab{b}})]{CP_LSTS_MONETARY_POLICY}
Alessandro Casini and Pierre Perron.
\newblock Change-point analysis of time series with evolutionary spectra.
\newblock \emph{Journal of Econometrics}, 242\penalty0 (2):\penalty0 105811, 2024{\natexlab{b}}.
\newblock ISSN 0304-4076.
\newblock \doi{https://doi.org/10.1016/j.jeconom.2024.105811}.

\bibitem[Chazal and Michel(2021)]{Chazal2021}
Fr{\'e}d{\'e}ric Chazal and Bertrand Michel.
\newblock An {{Introduction}} to {{Topological Data Analysis}}: {{Fundamental}} and {{Practical Aspects}} for {{Data Scientists}}.
\newblock \emph{Frontiers in Artificial Intelligence}, 4, 2021.
\newblock ISSN 2624-8212.

\bibitem[Dahlhaus(2000)]{Dahlhaus}
Rainer Dahlhaus.
\newblock A likelihood approximation for locally stationary processes.
\newblock \emph{Annals of Statistics}, 28:\penalty0 1762--1794, 2000.
\newblock \doi{https://doi.org/10.1214/aos/1015957480}.

\bibitem[Edelsbrunner and Harer(2008)]{TDA_EDELSBRUNNER_HARER}
Herbert Edelsbrunner and John Harer.
\newblock Persistent homology—a survey.
\newblock \emph{Discrete \& Computational Geometry}, 453:\penalty0 257--282, 2008.
\newblock \doi{10.1090/conm/453/08802}.

\bibitem[Edelsbrunner et~al.(2002)Edelsbrunner, Letscher, and Zomorodian]{TDA_EDELSBRUNNER}
Herbert Edelsbrunner, David Letscher, and Afra Zomorodian.
\newblock Topological persistence and simplification.
\newblock \emph{Discrete \& Computational Geometry}, 28:\penalty0 511–533, 2002.
\newblock \doi{doi.org/10.1007/s00454-002-2885-2}.

\bibitem[El-Yaagoubi et~al.(2023)El-Yaagoubi, Chung, and Ombao]{TDA_MULTIVARIATE_TS_ANASS}
Anass~B. El-Yaagoubi, Moo~K. Chung, and Hernando Ombao.
\newblock Topological data analysis for multivariate time series data.
\newblock \emph{Entropy}, 25:\penalty0 1509, 2023.
\newblock \doi{10.3390/e25111509}.

\bibitem[Enikeeva and Klopp(2021)]{enikeeva2021changepointdetectiondynamicnetworks}
Farida Enikeeva and Olga Klopp.
\newblock Change-point detection in dynamic networks with missing links, 2021.
\newblock URL \url{https://arxiv.org/abs/2106.14470}.

\bibitem[Fiecas and Ombao(2016)]{DYNAMIC_BRAIN_PROCESSES}
Mark Fiecas and Hernando Ombao.
\newblock Modeling the evolution of dynamic brain processes during an associative learning experiment.
\newblock \emph{Journal of the American Statistical Association}, 111:\penalty0 1440--1453, 2016.
\newblock \doi{10.1080/01621459.2016.1165683}.

\bibitem[Galadí et~al.(2021)Galadí, {Silva Pereira}, {Sanz Perl}, Kringelbach, Gayte, Laufs, Tagliazucchi, Langa, and Deco]{NON_STATIONARY_BRAIN}
J.A. Galadí, S.~{Silva Pereira}, Y.~{Sanz Perl}, M.L. Kringelbach, I.~Gayte, H.~Laufs, E.~Tagliazucchi, J.A. Langa, and G.~Deco.
\newblock Capturing the non-stationarity of whole-brain dynamics underlying human brain states.
\newblock \emph{NeuroImage}, 244:\penalty0 118551, 2021.
\newblock \doi{https://doi.org/10.1016/j.neuroimage.2021.118551}.

\bibitem[Gidea et~al.(2020)Gidea, Goldsmith, Katz, Roldan, and Shmalo]{CP_CRYPTOS}
Marian Gidea, Daniel Goldsmith, Yuri Katz, Pablo Roldan, and Yonah Shmalo.
\newblock Topological recognition of critical transitions in time series of cryptocurrencies.
\newblock \emph{Physica A: Statistical Mechanics and its Applications}, 548:\penalty0 123843, 2020.
\newblock \doi{https://doi.org/10.1016/j.physa.2019.123843}.

\bibitem[Henneuse(2024)]{henneuse2024}
Hugo Henneuse.
\newblock Persistence diagram estimation of multivariate piecewise h\"older-continuous signals, 2024.
\newblock URL \url{https://arxiv.org/abs/2403.19396}.

\bibitem[Islambekov et~al.(2019)Islambekov, Yuvaraj, and Gel]{CP_ENVIRONMENT}
Umar Islambekov, Monisha Yuvaraj, and Yulia~R. Gel.
\newblock Harnessing the power of topological data analysis to detect change points.
\newblock \emph{Environmetrics}, 31\penalty0 (2):\penalty0 e2612, December 2019.
\newblock \doi{10.1002/env.2612}.

\bibitem[Jenkins and Priestley(1957)]{PSD_Priestley_1957}
G.~M. Jenkins and M.~B. Priestley.
\newblock The spectral analysis of time-series.
\newblock \emph{Journal of the Royal Statistical Society. Series B (Methodological)}, 19\penalty0 (1):\penalty0 1--12, 1957.
\newblock URL \url{https://www.jstor.org/stable/2983992}.

\bibitem[Jiao et~al.(2023)Jiao, Frostig, and Ombao]{JIAO2023}
Shuhao Jiao, Ron~D. Frostig, and Hernando Ombao.
\newblock Break point detection for functional covariance.
\newblock \emph{Scandinavian Journal of Statistics}, 50\penalty0 (2):\penalty0 477--512, 2023.
\newblock \doi{https://doi.org/10.1111/sjos.12589}.

\bibitem[Lee et~al.(2007)Lee, Qiu, Yu, and Lin]{BEARING_DATA_SET}
J.~Lee, H.~Qiu, G.~Yu, and J.~Lin.
\newblock Bearing data set.
\newblock Technical report, Rexnord Technical Services, 2007.
\newblock IMS, University of Cincinnati.

\bibitem[Lee(1997)]{PeriodSmootherLee}
Thomas Lee.
\newblock A simple generalised crossvalidation method of span selection for periodogram smoothing.
\newblock \emph{Biometrika}, 84:\penalty0 965--969, 1997.
\newblock \doi{https://doi.org/10.1093/biomet/84.4.965}.

\bibitem[Mancho-Fora et~al.(2020)Mancho-Fora, Montalà-Flaquer, Farràs-Permanyer, Bartrés-Faz, Vaqué-Alcázar, Peró-Cebollero, and Guàrdia-Olmos]{Mancho2020}
Núria Mancho-Fora, Marc Montalà-Flaquer, Laia Farràs-Permanyer, David Bartrés-Faz, Lídia Vaqué-Alcázar, Maribel Peró-Cebollero, and Joan Guàrdia-Olmos.
\newblock Resting-state functional connectivity dynamics in healthy aging: An approach through network change point detection.
\newblock \emph{Brain Connect}, 10\penalty0 (3):\penalty0 134--142, 2020.
\newblock \doi{10.1089/brain.2019.0735}.

\bibitem[Ombao et~al.(2001)Ombao, Raz, Strawderman, and von Sachs]{PeriodSmootherGCV}
Hernando Ombao, Jonathan Raz, Robert Strawderman, and Rainer von Sachs.
\newblock A simple generalised crossvalidation method of span selection for periodogram smoothing.
\newblock \emph{Biometrika}, 88:\penalty0 1186--1192, 2001.
\newblock \doi{https://doi.org/10.1093/biomet/88.4.1186}.

\bibitem[Seversky et~al.(2016)Seversky, Davis, and Berger]{severskyTimeSeriesTopologicalData2016}
Lee~M. Seversky, Shelby Davis, and Matthew Berger.
\newblock On {{Time-Series Topological Data Analysis}}: {{New Data}} and {{Opportunities}}.
\newblock In \emph{2016 {{IEEE Conference}} on {{Computer Vision}} and {{Pattern Recognition Workshops}} ({{CVPRW}})}, pages 1014--1022, {Las Vegas, NV, USA}, June 2016. {IEEE}.
\newblock \doi{10.1109/CVPRW.2016.131}.

\bibitem[Truong et~al.(2020)Truong, Oudre, and Vayatis]{TRUONG2020107299}
Charles Truong, Laurent Oudre, and Nicolas Vayatis.
\newblock Selective review of offline change point detection methods.
\newblock \emph{Signal Processing}, 167:\penalty0 107299, 2020.
\newblock ISSN 0165-1684.
\newblock \doi{https://doi.org/10.1016/j.sigpro.2019.107299}.
\newblock URL \url{https://www.sciencedirect.com/science/article/pii/S0165168419303494}.

\bibitem[Zhang et~al.(2024)Zhang, Zhang, Sun, and Wang]{Zhang2024}
Yuzhao Zhang, Jingnan Zhang, Yifan Sun, and Junhui Wang.
\newblock Change point detection in dynamic networks via regularized tensor decomposition.
\newblock \emph{Journal of Computational and Graphical Statistics}, 33\penalty0 (2):\penalty0 515--524, 2024.
\newblock \doi{10.1080/10618600.2023.2240864}.
\newblock URL \url{https://doi.org/10.1080/10618600.2023.2240864}.

\end{thebibliography}

\end{document}